\documentclass[aps]{revtex}
\usepackage{epsfig}

\begin{document}
%
\draft
\title{Precision Pion--Proton Elastic Differential Cross Sections \\
  at Energies Spanning the $\Delta$ Resonance}
%
\author{M.M. Pavan\footnote[2]{Corresponding Author: e-mail -
    marcello.pavan@triumf.ca}\footnote{Present address: TRIUMF, Vancouver,
    British Columbia V6T 2A3}, J.T. Brack\footnote[3]{Present address:
    Department of Physics, University of Colorado, Boulder, Colorado
    80309}, F. Duncan\footnote[4]{Present address: Department of Physics,
    Queen's University, Kingston, Ontario K7L 3N6}, A.
  Feltham\footnote[5]{Present address: Broadcom Canada Ltd.\, Richmond,
    British Columbia V6V 2Z8}, G. Jones, J. Lange\footnote[6]{Present
    address: Defense Research Establishment,
    Ottawa, Ontario K1A 0Z4}, \\
  K.J. Raywood\footnotemark[1], M.E. Sevior\footnote[7]{Present address:
    School of Physics, University of Melbourne, Parkville, Victoria 3052,
    Australia } } \address{Department of Physics, University of British
  Columbia, Vancouver, British Columbia V6T-1Z1}

\author{R. Adams, D.F. Ottewell,
  G.R. Smith\footnote[9]{Present address: Jefferson Lab, 12000 Jefferson
    Avenue, MS 12H Newport News, Virginia 23606}, B. Wells,
  R.L. Helmer} \address{TRIUMF, Vancouver, British Columbia V6T-2A3}

\author{E.L. Mathie, R. Tacik} \address{University of Regina, Regina,
  Saskatchewan S4S-0A2}

\author{R.A. Ristinen} \address{Department of Physics, University of
  Colorado, Boulder, Colorado 80309}

\author{I.I. Strakovsky\footnote[10]{Present address: Center for Nuclear
    Studies and Department of Physics,
    The George Washington University, Washington, DC 20052} }
\address{Petersburg Nuclear Physics Institute, Gatchina, St. Petersburg,
    Russia, 188350}

\author{H-M. Staudenmaier} \address{Institute for Theoretical Particle
  Physics, University of Karlsruhe, Karlsruhe D-76128, Germany}

\date{\today} \maketitle
\begin{abstract}
  
  A precision measurement of absolute \mbox{$\pi^{\pm}p$}\/ elastic
  differential cross sections at incident pion laboratory kinetic energies
  from $T_{\pi}$\/ = 141.15 to 267.3 MeV is described.  Data were obtained
  detecting the scattered pion and recoil proton in coincidence at 12
  laboratory pion angles from 55$^{\circ}$ to 155$^{\circ}$ for
  \mbox{$\pi^{+}p$}\ , and six angles from 60$^0$ to 155$^0$ for
  \mbox{$\pi^{-}p$}\@. Single arm measurements were also obtained for
  $\pi^+$p energies up to 218.1 MeV, with the scattered $\pi^+$ detected at
  six angles from 20$^0$ to 70$^0$.  A flat--walled, super-cooled liquid
  hydrogen target as well as solid CH$_{2}$\ targets were used. The data
  are characterized by small uncertainties, $\sim$1-2\% statistical and
  $\sim$1-1.5\% normalization. The reliability of the cross section results
  was ensured by carrying out the measurements under a variety of
  experimental conditions to identify and quantify the sources of
  instrumental uncertainty.  Our lowest and highest energy data are
  consistent with overlapping results from TRIUMF and LAMPF\@. In general,
  the Virginia Polytechnic Institute SM95 partial wave analysis solution
  describes our data well, but the older Karlsruhe--Helsinki PWA solution
  KH80 does not.

\end{abstract}



\newpage
 
\section{Introduction}
\label{sec:intro}

The pion-nucleon ($\pi$N) system at energies up to the first ($\Delta$)
resonance continues to be an area of keen theoretical and experimental
interest.  This is due in large part to the intimate connection of
low energy $\pi$N scattering to SU(2) quantum chromodynamics (QCD) at low
energies. By studying the low energy interactions of pions and nucleons,
one is able to probe the confinement scale structure of QCD (via an
effective theory, chiral perturbation theory, or ChPT \cite{gas84,bkm95}).
Two key areas of interest in low energy $\pi$N QCD centre on determinations
of the precise values of the $\pi$N sigma term $\Sigma$
\cite{sai97,cheng71,gas91} and the $\pi$NN coupling constant $f_{\pi
  NN}^{2}$ \cite{sai97,deSw97}.  The $\pi$N sigma term is fundamental to
low energy QCD since it quantifies the explicit breaking of chiral symmetry
due to the non-zero up and down quark masses.  The coupling constant $f^2$
is {\it the}\/ fundamental free parameter in ChPT involving nucleons
\cite{gas88}. It also appears in the well-known Goldberger--Treiman
relation \cite{gol58,pag75}, which relates $f^{2}$ to the accurately known
pion decay constant F$_{\pi}$, nucleon mass M, and axial-vector coupling
constant, g$_{A}$. The analogous Dashen-Weinstein sum rule
\cite{pag75,das69,dom85} relates $f^2$ to coupling constants involving
kaons, sigma and lambda baryons, and is closely connected to the quark
condensate $\overline{q}q$ \cite{fuc94,goit99}.

Despite many years of investigation, there is still no broad
consensus regarding the precise values of these important quantities. The
``sigma term puzzle" \cite{jaf87} refers to the historical discrepancy
between the phenomenologically determined value \cite{koch82} and the
theoretical prediction \cite{gas81}, a discrepancy which could imply a
large strange quark content of the proton. The puzzle has yet to be
resolved \cite{sai97,gas91}. The value of the coupling constant $f^2$ has
been controversial as well \cite{deSw97}, with recent results split roughly
into two groups: $f^2 \sim 0.0795$ \cite{kh80,eric95} and 0.0755
\cite{deSw97,sm95,tim97}.  The $\sim$5\% difference has significant
implications for the aforementioned Goldberger-Treiman and Dashen-Weinstein
relations, as well as for any model employing the $\pi$NN vertex (e.g.\ the
Bonn NN potential \cite{mach91,mach94}).

\begin{figure}[hp]
  \begin{tabular}{l@{\hspace{4mm}}r}
    \epsfig{file=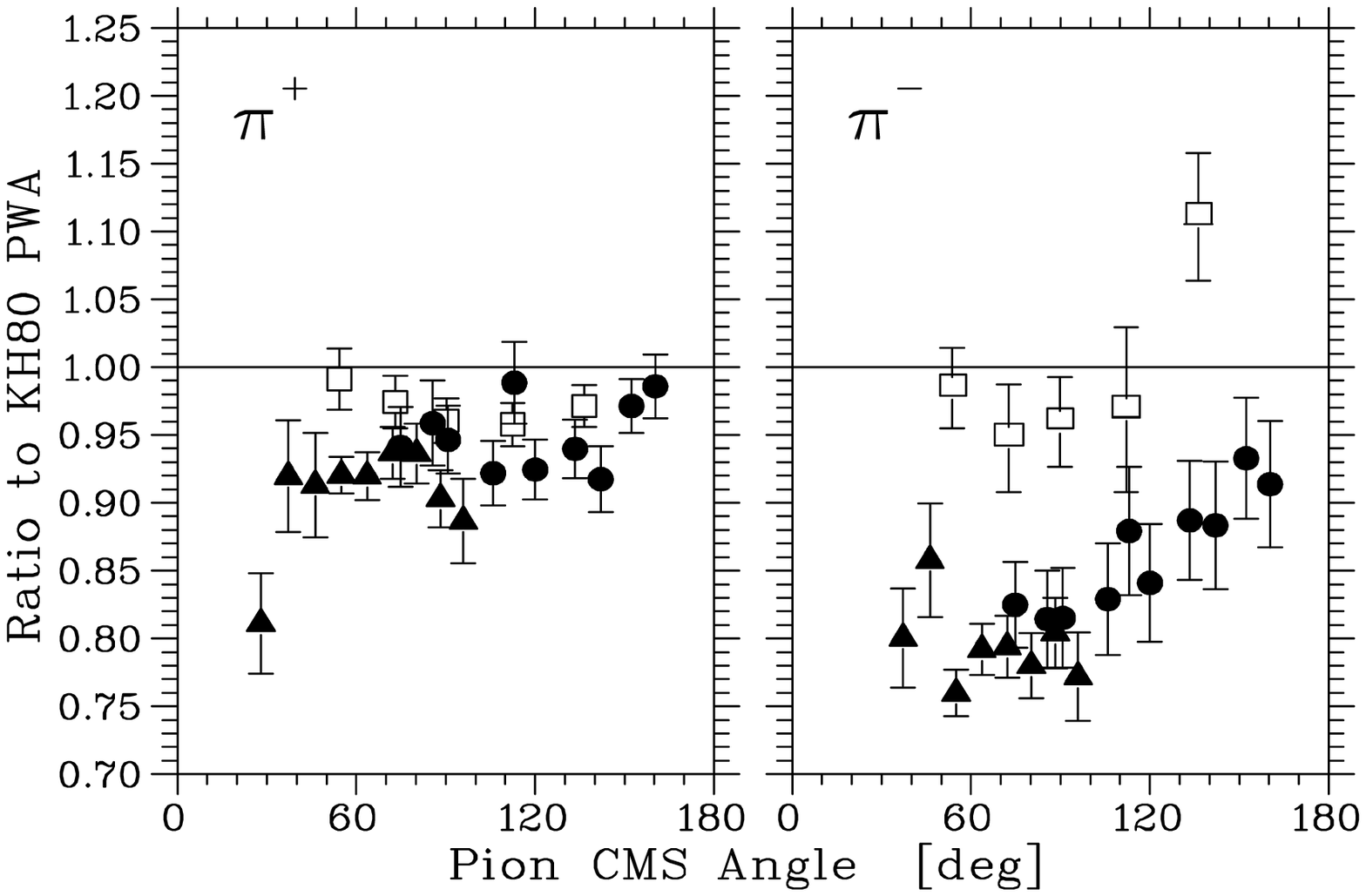,width=8.4cm} &
    \epsfig{file=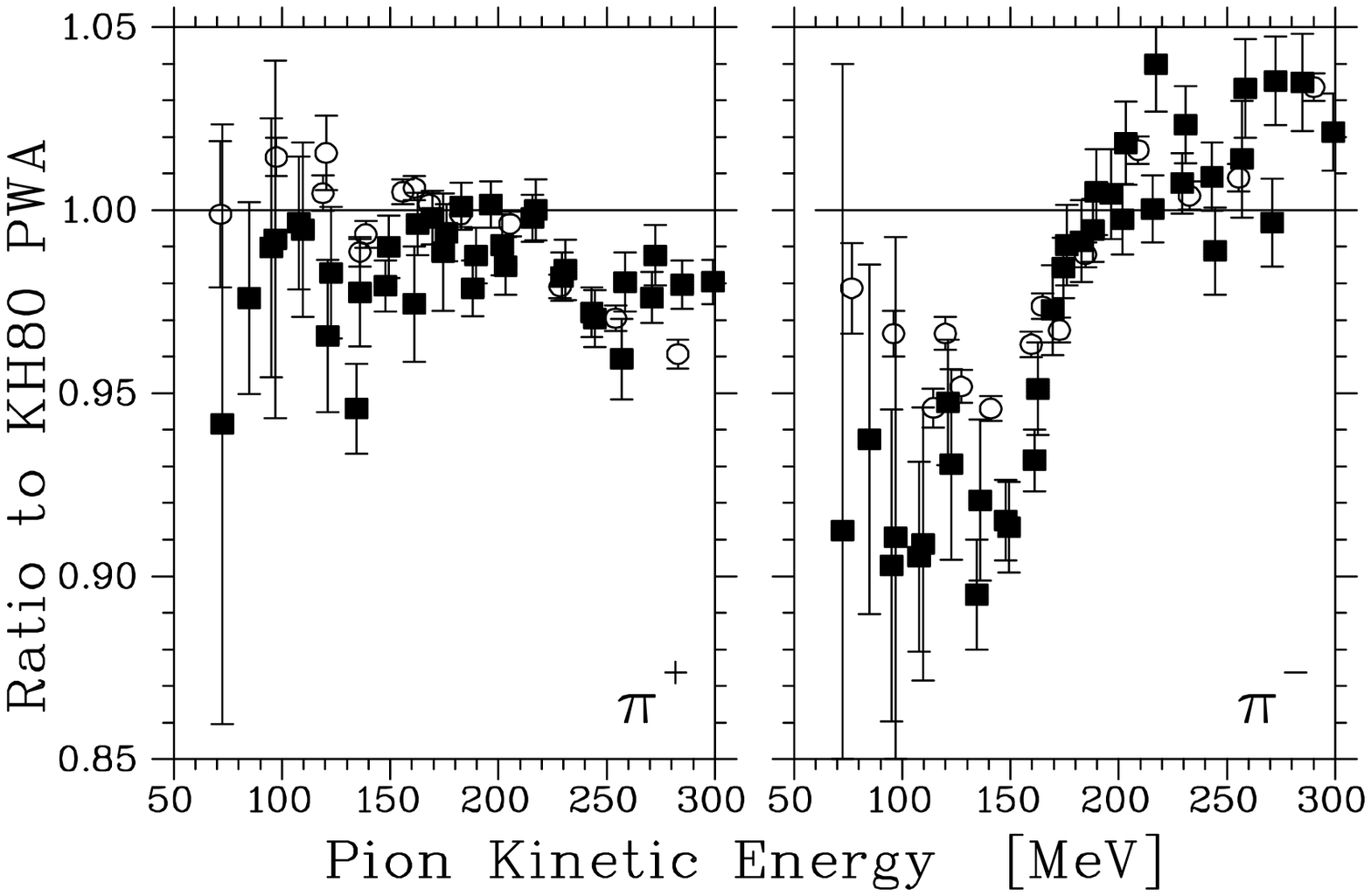,width=8.4cm} 
  \end{tabular}
  \caption{
    {\bf Left:} Differential cross sections near T$_{\pi}=$117 MeV of
    Bussey et al.~\protect\cite{bus73} (open) versus Brack et
    al\protect\cite{bra86,bra95} (solid) plotted as a ratio to the KH80 PWA
    solution\protect\cite{kh80} at their respective energies.  {\bf Right:}
    Total cross sections as a function of energy showing the results of
    Carter et al.  \protect\cite{car71} (open) versus Pedroni et
    al.\protect\cite{ped78} (solid).}
  \label{fig:compareBuggData}
\end{figure}

A major reason for the difficulty in determining $\Sigma$ and $f^2$ arises
from historical incompatibilities in the $\pi$N scattering database
\cite{pinNewsLett}. As the determination of these parameters requires
extrapolations of the scattering amplitudes to nonphysical kinematic points,
an internally consistent database of precision data is crucial for reliable
results.  The most trustworthy analyses employ $\pi$N dispersion relations
\cite{bible}. Since the $P_{33}$ $\pi$N partial wave amplitude in the delta
resonance ($\Delta$) region dominates the dispersion relations used to
obtain $\Sigma$ and $f^2$\cite{bible}, it is crucial that the data in this
energy region be reliable, mutually consistent, and of high quality.
Differential cross section data are of particular importance, yet to date
only one comprehensive data set exists spanning the $\Delta$
resonance\footnote{Other data sets exists in this region, e.g.\ 
  \protect\cite{otherData}, but the data are much more limited in number.},
the work of Bussey et al. \cite{bus73}. Unfortunately, the Bussey data, and
that of the companion total cross section work of Carter et al.
\cite{car71}, are generally at variance with partial-wave analyses
\cite{sm95,tim97} based on recent differential cross section data below 140
MeV kinetic energy \cite{bra86,bra95} and above 267 MeV \cite{sad87}, as
well as with the other total cross section data of Pedroni et al.
\cite{ped78} across the resonance (Fig.~\ref{fig:compareBuggData}).
Moreover, the normalization uncertainties in the Bussey et al.\ data
recently have been increased {\em post priori} \cite{bugg-renorm},
indicating possible problems with the data, thus making a new measurement
all the more relevant and important.

The goal of the work described in this paper was to provide a new
comprehensive set of precision $\pi^{\pm}p$ absolute differential cross
section data characterized by reliable estimates of systematic
uncertainties at energies spanning the $\Delta$ resonance. Experimental
details such as the apparatus, the data acquisition system, and the
data-taking techniques are described in Sec.~\ref{sec:expt}.  The offline
data analysis and the Monte Carlo simulations are detailed in
Sec.~\ref{sec:data_anal}, and the results are presented in
Sec.~\ref{sec:results}, followed by a discussion in
Sec.~\ref{sec:conclusion}. Additional details can be found elsewhere
\cite{mythesis}.

\section{Experiment}
\label{sec:expt}

The experiment was conducted on the M11 pion channel at TRIUMF\@.  Data
were obtained for both \mbox{$\pi^{+}$p}\/ and \mbox{$\pi^{-}$p}\/ elastic
scattering at incident pion lab kinetic energies of $T_{\pi}$\/ =
141.15$\pm$0.6, 168.8$\pm$0.7, 193.2$\pm$0.7, 218.1$\pm$0.8, 240.9$\pm$0.9,
and 267.3$\pm$0.9 MeV, and for $\pi^+$p, at 154.6$\pm$0.6 MeV as well.
These energies were chosen to span the $\Delta$ resonance, to overlap the
highest energy used by Brack et al. \cite{bra86} and the lowest of Sadler
et al. \cite{sad87}, and to coincide with those of the \mbox{$\pi^{\pm}$p}\ 
analyzing power measurements of Sevior et al. \cite{sev88} since the
availability of both differential cross sections and analyzing powers at
the same energies facilitates single-energy partial wave analyses.  For all
energies, \mbox{$\pi^{\pm}$p}\/ two--arm coincidence data were obtained at
middle and near-backward angles, while single-arm results (with only the
scattered $\pi^+$ detected) were obtained at near--forward angles at
141.15, 168.8, and 218.1 MeV.

Although the main goal of the experiment was to obtain statistically
precise results, a study of the various sources of systematic uncertainty
was also an important feature of this work.  As pointed out by Bugg
\cite{pin2-bugg}, the six elements essential to a measurement of absolute
differential cross sections are accurate knowledge of the: 1) beam
intensity, 2) beam composition, 3) beam momentum, 4) target thickness, 5)
solid angles, and 6) backgrounds.  To ensure confidence in the results, the
uncertainties claimed for the measurements should be based on the extent to
which the values at each fixed kinematical point are independent of
measured variations in the experimental conditions.  In this way systematic
uncertainties can be more accurately and reliably determined.
 
\begin{figure}[hb]
\begin{center}
  \epsfig{file=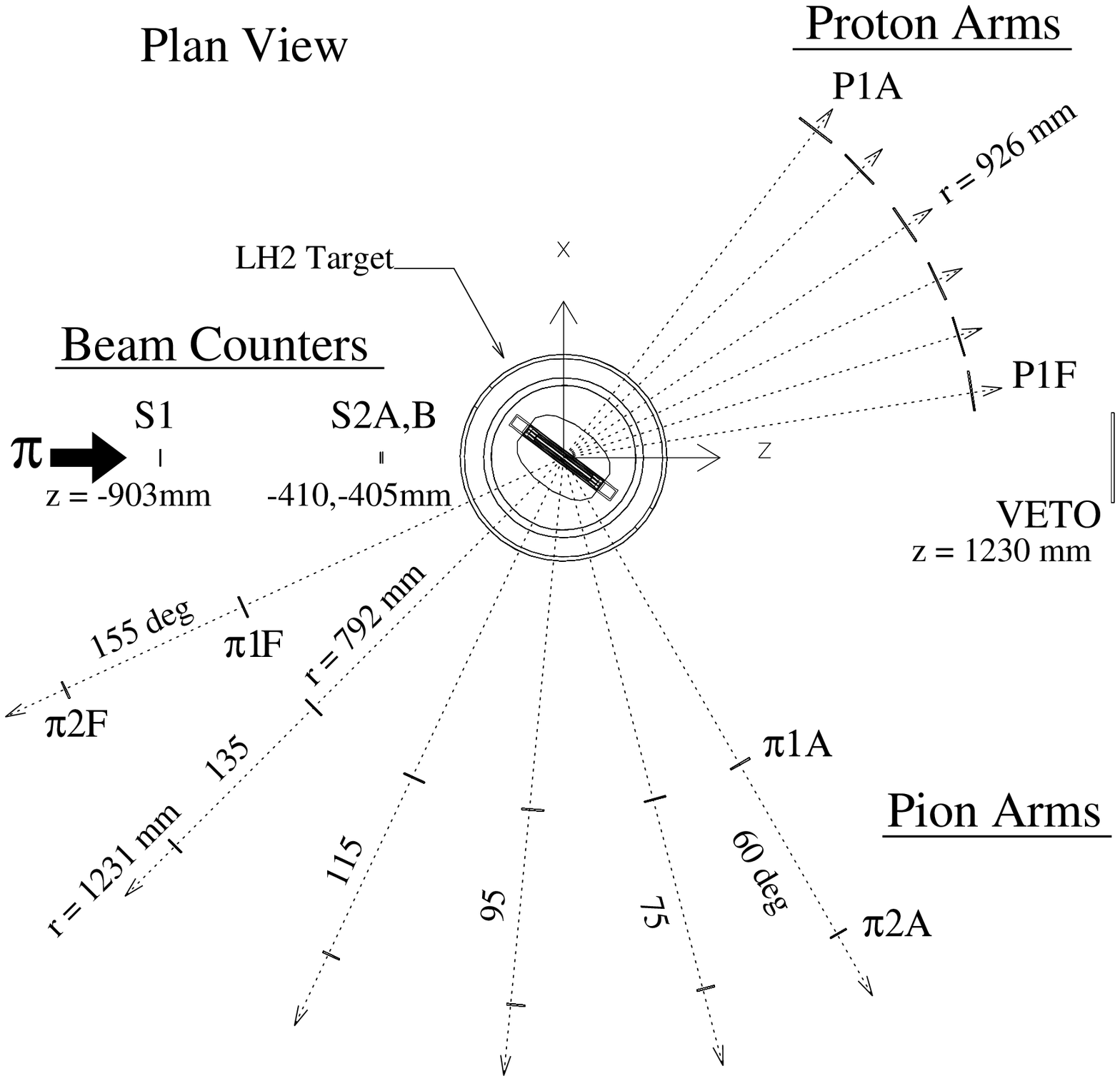,width=8.6cm}
\end{center}
\caption{
  Plan view of the TOF spectrometer, showing the distances between
  centrelines of the counters and target, and the pion arm angles for `Set
  A'. Proton arm angles vary with energy (see text). The pion beam enters
  at centre left.}
\label{fig:geantFig1}
\end{figure}

The general layout of the experiment is illustrated in
Fig.~\ref{fig:geantFig1}.  A complete description of the detector elements
is presented in the following sections.

\subsection{Pion Beam}
\label{sec:beam}

For our experiment, the pion beam originated at a beryllium target in the
140 $\mu$A primary proton line, BL1A, which yields a proton beam consisting
of pulses $\sim$3-4 ns wide occurring with a repetition rate of 23.06 MHz.
After momentum selection in the M11 pion channel, the pions were brought to
a doubly-achromatic double-focus at the target location.

The incoming pion beam was detected by three beam-defining scintillators
(S1, S2A, S2B) operating in three-fold coincidence and placed upstream of
the target. All consisted of 1.59 mm thick NE110, wrapped by a single layer
of 0.025 mm aluminum foil and 0.263 mm electrical tape.  These counters
were connected by short straight lucite light guides to photomultiplier
tubes mounted on high-rate transistorized bases \cite{reginatubes}.
Alignment of the counters was carried out using an optical transit.  The
25.4 mm wide by 102 mm high S1 counter was placed 903 mm
upstream\footnote{All distances are between centres, unless otherwise
  stated} of the target centre, and 187 mm from the exit of the 200 mm
diameter M11 beam pipe.  The 12.7 mm wide by 44.5 mm high S2A and S2B
counters were placed 410 and 405 mm, respectively, upstream of the target
location. They were mounted so that the S2A phototube was above the
scintillator and the S2B phototube below. This ensures that muons from pion
decay downstream of S1 would not cause erroneous coincidences by producing
$\check{\text{C}}$erenkov light in a S2 light guide that could be detected
by the phototube, since such muons could hit only one of the two S2 light
guides, and so such events were eliminated by the coincidence requirement.
Although the exact spot sizes depended on the settings of the rate-defining
aperture (jaws) at the front end of the channel, the beam distributions at
the target were typically 10 x 8 mm$^2$ and 1$^0$ x 4$^0$ at half-maximum.

Two counter telescopes, each with two scintillators in coincidence, were
used to monitor beam intensity relative to the beam counters.  One set
above the M11 beam pipe exit was used to detect muons from pion decay in
the channel.  The other set was mounted at beam height in the experimental
area and oriented to view particles back-scattered from the S2A,B counters.

To determine the fraction of beam bursts containing only one pion (see
Sec.~\ref{sec:multpion}), it was necessary to know the full beam rate on
target, which was somewhat larger than the rate measured by the beam
counters.  For this purpose, a 201 x 201 x 6.35 mm$^3$ VETO paddle was
placed 1230 mm downstream of the target position.  This counter intercepted
$>$95\% of the incoming pion beam.

Each beam counter signal was electronically fanned--out and fed to a
constant--fraction--discriminator (CFD). The counter voltages and CFD
thresholds were set so that all minimum ionizing particles were detected
but the thermionic tube noise was not.  A second signal from the S2B
counter was fed to a leading-edge discriminator in which the threshold was
set to detect only very large pulse height signals (S2BH), corresponding to
the proton contamination in the incident beam during $\pi^+$ running.
Inversion of the S2BH output ($\overline{\mbox{S2BH}}$) thus indicated a
``no--proton'' event.  Use of this signal together with a differential
absorber at the channel mid-plane reduced proton contamination in the beam
definition to $<$0.1\%.
   
An incident particle was identified electronically by the four--fold
coincidence $\mbox{BEAM} \equiv
\mbox{S1}\cdot\mbox{S2A}\cdot\mbox{S2B}\cdot \overline{\mbox{S2BH}}$. The
logic signal S2B defined the timing for the entire system.  The tight
angular definition of this telescope of beam counters ensured that all BEAM
coincidences corresponded to a particle at the target (except for those
pions which decayed or suffered hadronic interaction prior to reaching the
target, as described in Sec.~\ref{sec:beamloss}). Particle identification
(see Sec.~\ref{sec:beamcomp}) was realized by measuring the relative
times-of-flight (TOF) of the particles down the pion channel, values
obtained from the time differences between the BEAM coincidence and the
TCAP signal, the latter a signal produced by a capacitive pickup in the
primary proton beam line.

In order to monitor the incident beam, a special ``beam samples'' trigger
(SAMPLE) was constructed which utilized only BEAM coincidences selected
randomly by an adjustable clock pulse. Since the vast majority of BEAM
triggers did not cause \mbox{$\pi$p}\/ events, this trigger provided an
unbiased sample of events striking the beam counters.

In order to assess the fraction of BEAM coincidences consisting of more
than one pion, pions detected in the two buckets following the one that
triggered the spectrometer were also monitored. Circuits were constructed
to detect BEAM hits in 2 and 3 consecutive beam buckets after an initial
BEAM event. According to Poisson statistics, the probability of at least
one hit occurring in each of `m' consecutive beam buckets is
$(1-\text{e}^{-\lambda})^{m}$, where $\lambda$ is the probability of a pion
occurring in a single beam bucket.  This relationship was found to be well
reproduced throughout the experiment. As discussed in
Sec.~\ref{sec:multpion}, such information was required to correct the beam
rate for those events characterized by more than one pion in a beam bucket.

For all of our $\pi^{+}$\/ measurements, the channel slit width was set at
18 mm, corresponding to a 1\% FWHM $\delta$p/p momentum spread
\cite{triumfUserMan}. For the $\pi^{-}$\/ runs where the fluxes were lower,
the momentum spread was set at 2\%, except for 267 MeV, where it was 2.5\%.
The channel jaws were adjusted at each energy to provide typically 1.5MHz
and 2MHz target rates for $\pi^{+}$\/ and $\pi^{-}$, respectively.

A comprehensive beam tuning and calibration program \cite{mythesis} was
undertaken immediately prior to the experiment in order to gain a detailed
understanding of the pion beam characteristics.  Two issues which arose
from those studies deserve particular mention.  The spot size and
divergence of the beam at our target location were found to vary slightly
with the aperture of the front--end rate restricting jaws in the channel.
Consequently, the values of the jaw apertures were recorded for all
data-taking runs, since knowledge of the beam size and divergence was
required for accurate modelling by the Monte Carlo simulation programs (see
Sec.~\ref{sec:data_anal} and and Appendix~\ref{appdx:simulDetails}).  It
was also found that although the pion beam trajectory and size were rather
insensitive to the horizontal position of the primary proton beam on the
beryllium pion production target, they were somewhat sensitive to the
vertical position. To monitor beam movement, a square, four-paddle
hodoscope centred on the beam was placed 2480 mm downstream of the target
location.  The rate on each hodoscope paddle was continuously monitored and
written to the data acquisition stream throughout the experiment.

\subsubsection{Beam Momentum}
\label{sec:momencal}

\begin{figure}
\begin{center}
  \mbox{\epsfig{file=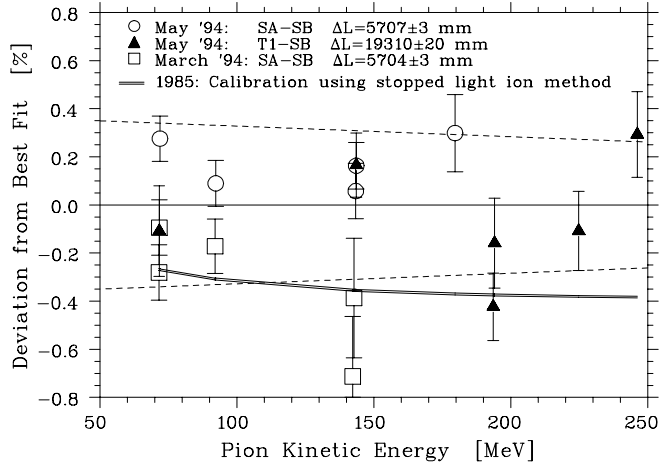,width=8.6cm}}
\end{center}
\caption{
  Results from our pion channel energy calibration using the $\pi$-e TOF
  difference technique. The data points show the \% deviation in kinetic
  energy from the resulting best-fit calibration.  The dashed lines
  represent the estimated $\pm$0.2\% momentum uncertainty. Also shown is
  the calibration from the 1985 analysis \protect\cite{m11-designnote},
  which used the technique of measuring the energy of light ions in the
  beam when stopped in a silicon counter.
  }
\label{fig:m11-calib}
\end{figure}

As the central pion momentum transmitted by the channel is linearly related
to the magnetic field strength of the first channel dipole (B1) measured by
an NMR probe set at the magnet mid-plane, momentum calibrations were
carried out during the tuning phase of the experiment and again near its
completion, using the traditional technique (SSBD) \cite{m11-designnote} of
stopping light ions produced at the production target in a silicon counter
in vacuum at the beam pipe exit. However, after the experiment, we were
made aware \cite{m11-joramhelp} of a pulse-height defect issue
\cite{martini,kemper,langley,haynes,coche,king} which rendered these
results unreliable. Consequently, another detailed calibration was
performed subsequent to the experiment by measuring pion--electron TOF
differences between scintillators contained within an evacuated beam pipe
in the experimental area, and also between the TCAP signal from the protons
in the primary beam line and a scintillator in the experimental area. The
technique exploits the fact that the electrons travel at essentially the
speed of light and so provide an absolute velocity scale. Details are
provided in Ref.~\cite{mythesis}. Data from these measurements
(Fig.~\ref{fig:m11-calib}) yielded the M11 channel momentum calibration:
\begin{equation}
\label{eqn:p-calib}
\mbox{P}_{\mbox{M11}} [\mbox{MeV}/c] = 326.7\cdot(\mbox{B1}-0.00171)\pm 0.2\% 
\end{equation}
where B1 is the magnetic field strength in Tesla, and the $\pm$0.2\%
uncertainty in $\mbox{P}_{\mbox{M11}}$ corresponds to the spread in the
calibration points from the best fit line shown in
Fig.~\ref{fig:m11-calib}. The previously-accepted M11
calibration~\cite{m11-designnote} (which employed the SSBD method) is shown
in Fig.~\ref{fig:m11-calib} as well.  The $\approx$0.25\% discrepancy in
momentum between this calibration and the new one is consistent with the
size of the pulse-height defect effects discussed in
Refs.~\cite{martini,kemper,langley,haynes,coche,king}\footnote{As an
  uncertainty in the M11 calibration of about $\pm$0.5 MeV was quoted by
  Brack {\it et al}.\ \cite{bra86} for their differential cross section
  results at energies up to 140 MeV, their old calibration is within the
  0.2\% momentum uncertainty of the new one at 140 MeV.}.

For each run in the experiment, the energy loss through the mid-plane
absorber, the beam pipe exit window, and the in--beam counters (including
tape), air, target windows, etc.\ (for the LH$_{2}$\ target), and half the
target material at the appropriate angle were determined using the full
Bethe--Bloch equation \cite{lossprog}.  The uncertainty in this energy loss
was estimated as 10\% of the total loss ($\Delta$T$_{\pi}$ typically 2
MeV), and was added in quadrature to the $\pm$0.20\% momentum uncertainty
calibration to give the total uncertainty.  Although the pion beam energy
was fine-tuned for each different target configuration to give the desired
energy at the target centre, the energies were {\em not}\/ similarly
adjusted for the corresponding background runs, since the backgrounds were
small as was the energy-dependence of the background itself. The variations
in the cross sections associated with the resulting momentum uncertainty
are less than about 1.7\% for both our $\pi^{+}$p\/ and $\pi^{-}$p data.

\subsection{Time-of-Flight Spectrometer} 
\label{sec:tofspect}

The TOF spectrometer shown in Fig.~\ref{fig:geantFig1} consisted of the
beam monitoring scintillators, six arms consisting of pairs of thin
scintillation counters to detect the scattered pions, and six conjugate
arms of thin scintillation counters to detect the recoil protons during
coincidence running. Use of thin transmission scintillators ensured
virtually 100\% detection efficiency with negligible edge effects for both
$\pi^{+}$\/ and $\pi^{-}$\/ as well as protons\footnote{The pion detection
  efficiency of the lucite light guides attached to the scintillators was
  measured to be negligible.}. A pion arm consisted of a two--counter
telescope viewing the target, with each telescope comprised of two 3.2 mm
thick NE102 scintillators wrapped by a single layer of 0.025 mm aluminum
foil and 0.26 mm polyvinylchloride electrical tape.  The scintillators were
attached to the phototubes via lucite light guides. The
telescopes were bolted onto a machined table, with both scintillators of
each arm positioned using a transit located at the target centre, enabling
the angular positions to be known to better than $\pm$0.2$^0$.  The solid
angle defining `$\pi$2' counters were on average 40.03$\pm$0.06 mm wide by
99.90$\pm$0.09 mm high \cite{brack-pc} and were mounted 1231$\pm$3 mm from
the target centre.  The `$\pi$1' counters were 49 mm by 165 mm, and
situated 792 mm from the target centre. These dimensions and separations
were chosen in order to define a projected spot size of $\approx$60 mm
horizontal, $\approx$200 mm vertical at the target, large enough to cover
the whole interaction region while not severely restricting the acceptance
to muons arising from decay of scattered pions.

For those runs involving coincidence detection (\mbox{$\pi$p})\ of both
pions and protons, a set of six 90 mm by 400 mm by 3.2 mm thick (`P1')
scintillators were used as the recoil--proton detection arms.  These
scintillators were viewed from both the top and bottom by phototubes
coupled to lucite light guides bent at 90$^0$ so that the phototubes pointed
radially.  The scintillators were situated 926$\pm$3 mm from the target.
The base plates for these counters were positioned also on a machined table
using a transit, with slight adjustments provided in order for the proton
counters to be moved after every energy change to the angles conjugate to
the scattered pions. An accuracy of about $\pm$0.1$^0$ was achieved in the
angular positions.

\subsubsection{Two--Arm $\pi$p Coincidence Detection}
\label{sec:twoarmcoinc}
  
Pions scattered into the pion arms were identified by a $\pi$1$\cdot\pi$2
coincidence between counters in the same arm, in any one of the six
telescopes (i.e. $\Pi_{i} \equiv \pi\mbox{1}_{i} \cdot \pi\mbox{2}_{i}$).
Phototube voltages and discriminator thresholds were set just above the
noise signals and at about 35\% of the smallest pion pulses to ensure that
no good pion events would be lost.  Particles were identified by their TOF
to the $\pi$2 counters relative to the BEAM signal. Although neither the
timing nor pulse height information from $\pi$2 could distinguish pions
from those muons arising from pion decay between the target and $\pi$2,
this small muon contribution could be accurately accounted for by Monte
Carlo.  The $\pi$1 counters were positioned such that pions passing through
them would not strike the $\pi$2 light guide near the phototube junction,
which would enable forward--going $\check{\text{C}}$erenkov radiation to be
detected, but nevertheless a check was made with the $\pi$1 counters out of
the EVENT coincidence. In this case the false events produced by the
$\check{\text{C}}$erenkov radiation were easily discriminated against with
a timing cut, since the the light arising from true events hitting the
$\pi$2 scintillator counters had a longer path length to traverse before
reaching the phototube.

Proton arm events were signalled by a coincidence between the up and
down tubes of each of the P1 counters.  Logic signals obtained by
discriminating with CFDs were then fed to a meantimer to
establish the timing gate.  Prior to performing the actual experiment, each
counter was placed in the beam, and the phototube voltages and
discriminator thresholds were adjusted to cut half--way into the (minimum
ionizing) electron signal, thus ensuring that all protons were detected.
Candidate \mbox{$\pi$p}\/ scattering events (ARM) were identified by the
coincidence of BEAM with the coincidence output of signals from a pion arm
and its conjugate proton arm (i.e., $ \mbox{ARM}_{i} \equiv \mbox{BEAM}
\cdot \Pi_{i} \cdot \mbox{P}_{i} $).  The timing was set such that only
relatively fast particles in the pion arm and relatively slow particles in
the proton arm would satisfy the $\Pi \cdot $P coincidence.
  
\begin{figure}
\begin{center}
  \mbox{\epsfig{file=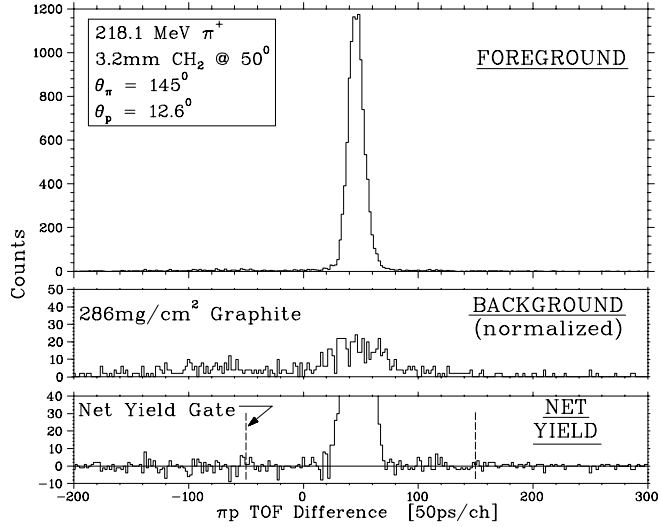, width=8.6cm}}
\end{center}
\caption{Foreground, normalized background, and net $\pi$p TOF
  difference spectrum at 145$^0$ laboratory angle for 218 MeV $\pi^{+}$p on
  a 3.2 mm CH$_{2}$\ target.  This setting had the 
  {\em worst}\/ background--to--foreground ratio ($\approx$7\%) of 
  {\em all}\/  the two--arm runs in the experiment (which were typically
  $\approx$1-3\%). A software yield--defining gate is shown by the vertical
  dashed lines.  \normalsize }
\label{fig:fgbgnet}
\end{figure}

The \mbox{$\pi$p}\/ scattering yield was obtained from the spectra of TOF
differences of particles to the pion counters relative to those to the
proton counters. The tight geometry of the counter pairs greatly suppressed
the dominant 3--body quasi--elastic
$\pi^{\pm}\mbox{A}\rightarrow\pi^{\pm}\mbox{p}\mbox{X}$ background, and,
combined with the timing requirement, also suppressed the quasi--free
absorption $\pi^{+}\mbox{A}\rightarrow\mbox{p}\mbox{p}\mbox{X}$ background.
These backgrounds for the two--arm coincidence measurements {\em never}\/
exceeded 7\% of the foreground at any angle or energy.
Figure~\ref{fig:fgbgnet} shows the yield spectrum for the {\em worst
  case}\/.

With the system set up as described, data were obtained for both $\pi^+$
and $\pi^-$ for the pion lab angles {60, 75, 95, 115, 135, 155} (``set A'')
at all energies with $\theta_{\text{tgt}}=53.6^{\circ}$, and an additional
set at {55, 65, 85, 105, 125, 145} (``set B'') for a few $\pi^+$p energies
with $\theta_{\text{tgt}}=50.6^{\circ}$.  The proton angles were adjusted
at each beam energy to the appropriate values conjugate to those of the
pion arms.

The possibility of an ARM$_i$ coincidence being generated by the detection
of a proton in the pion arm and a pion in the proton arm was completely
eliminated in all but a single case by the tight kinematical constraints
imposed by the pion--proton counter pairs.  The one case where such events
could occur was in the ``Set B'' configuration where both the pion and
proton angles were $\approx$55$^0$.  These events were easily separated
from the true \mbox{$\pi$p}\/ events by the TOF timing difference, and
therefore did not present a problem in the analysis.
  
As shown in Fig.~\ref{fig:geantFig1}, the targets were arranged such that
the pion arms faced the upstream surface of the target (with respect to the
incident beam), whereas the proton arms faced the downstream surface to
minimize proton energy loss and multiple scattering. The target angles were
chosen to minimize the target thickness for the lowest energy protons. The
requirement that these protons not suffer excessive energy loss and
multiple scattering on the way to the P1 counter limited the proton angle
to a maximum of about 55$^0$ at the lowest energy (141 MeV), corresponding
to a minimum pion angle of $\approx$55$^0$. For most runs (``set A''), the
forward--most pion arm was set at 60$^0$, corresponding to a P1 counter
angle of 53$^0$ at 141 MeV. The backward--most pion angle was limited to
155$^0$ by the requirement that the corresponding proton counter angle at
267 MeV (8.6$^0$) be situated safely outside the cone of the incident beam.

\subsubsection{Forward Angle Single--Arm Pion Detection}
\label{sec:onearmcoinc}

For pion angles less than about 50$^0$, the corresponding proton energies
were not large enough for the protons to escape from the liquid hydrogen
target (described in the following section).  Consequently, a set of
$\pi^{+}$\/ runs at forward pion angles were undertaken at
$\theta_{\pi}\in$ \{20, 30, 40, 50, 60, 70\} degrees with the proton arms
removed from the EVENT coincidence.  In this case an ARM$_i$ event was
defined by the coincidence BEAM$\cdot\Pi_i$
  
Candidate \mbox{$\pi$p}\/ events were identified by the TOF to the $\pi$2
counter.  In these single-arm liquid hydrogen target runs, the
foreground-to-background ratios were considerably poorer than in
coincidence mode, ranging from about 1.5:1 at 20$^0$, to about 7:1 at
70$^0$. A sample spectrum is illustrated in Fig.~\ref{fig:yield_1arm}.  The
reactions which contributed the bulk of the \mbox{$\pi^{+}$p}\/ single-arm
background included: pion elastic scattering from the mylar windows and
domes in the target (carbon, oxygen, and hydrogen), pion quasi--elastic
scattering from these materials (mainly carbon), and $\pi^{+}$\/ absorption
on quasi--deuterons in these same nuclei, producing two fast protons which
could satisfy the $\Pi$ timing gate.  However, despite the sizable
backgrounds, reliable background subtraction was possible.  The single-arm
runs were set up with the target oriented at -39.4$^{\circ}$ so that the
downstream window faced the middle pion arm.  The pion angles were chosen
to fill in the angles not already covered by the two-arm coincidence runs,
with some overlap to provide a consistency check.  The forward-most angle
was limited by the requirement that muons arising from decay of beam pions
would not cause a pion arm coincidence.

\subsection{Targets}
\label{sec:lh2target}
\label{sec:targets}
\label{sec:ch2target}

Since the use of solid targets in \mbox{$\pi$p}\/ elastic scattering
experiments has been the subject of some criticism \cite{pin2-bugg},
the coincidence measurements were taken with both thin
solid CH$_{2}$ and a novel flat-walled, super-cooled liquid hydrogen
(LH$_{2}$) target in order to lay this concern to rest. Although most of
the measurements described in this paper were done using the LH$_2$ target,
several measurements were repeated using the solid targets as a check on
systematic uncertainties.
 
The solid CH$_2$ targets consisted of 127 x 127 mm$^2$ square slabs of
$\rho$=0.93 gm/cm$^3$ CH$_{2}$, with a slab of 100 x 100 mm$^2$ square
carbon graphite used for background measurements.  The targets used and
their respective thicknesses are shown in Table~\ref{tab:target_thick}.
The densities were obtained from measurements of the linear dimensions
together with weights measured using a Mettler balance. The uniformity of
the linear thicknesses was checked using a machinist's comparator,
specified to be accurate to 2.5x10$^{-5}$ mm \cite{brack-thesis}. The
hydrogen and carbon contents of the CH$_{2}$\/ targets were determined to
1\% accuracy by chemical analyses provided by a commercial
laboratory~\cite{brack-thesis}. The stopping power for pions and protons in
the graphite background target was midway between those of the CH$_{2}$\/
`D' and `E' targets, the solid targets which were most often used in the
experiment.  Incidentally, these were the same targets used in the
experiments of Brack et al.~\cite{bra86,bra88}.
 
The targets were supported by thin aluminum frames attached to an aluminum
support bracket, and the whole assembly was mounted onto a machinist's
rotating table to provide accurate and reproducible angular adjustment. A
transit mounted downstream of the target position was used to check the
90$^0$ orientation (`edge on') of the target after every change or
adjustment of the target angle.  The 0$^0$ orientation was set by attaching
a mirror to the target, and then shining a He-Ne laser through the transit
viewpiece with the reflected light required to project back onto the laser
exit aperture.  In this way, the target angles were determined to
$\pm$0.25$^0$ (68\% confidence). The CH$_2$ target angle was fixed at
53.0$^{\circ}$ at all energies except 218.1 MeV, where it was
50.0$^{\circ}$.
 
\begin{figure}
  \begin{tabular}{l@{\hspace{4mm}}r}
  \epsfig{file=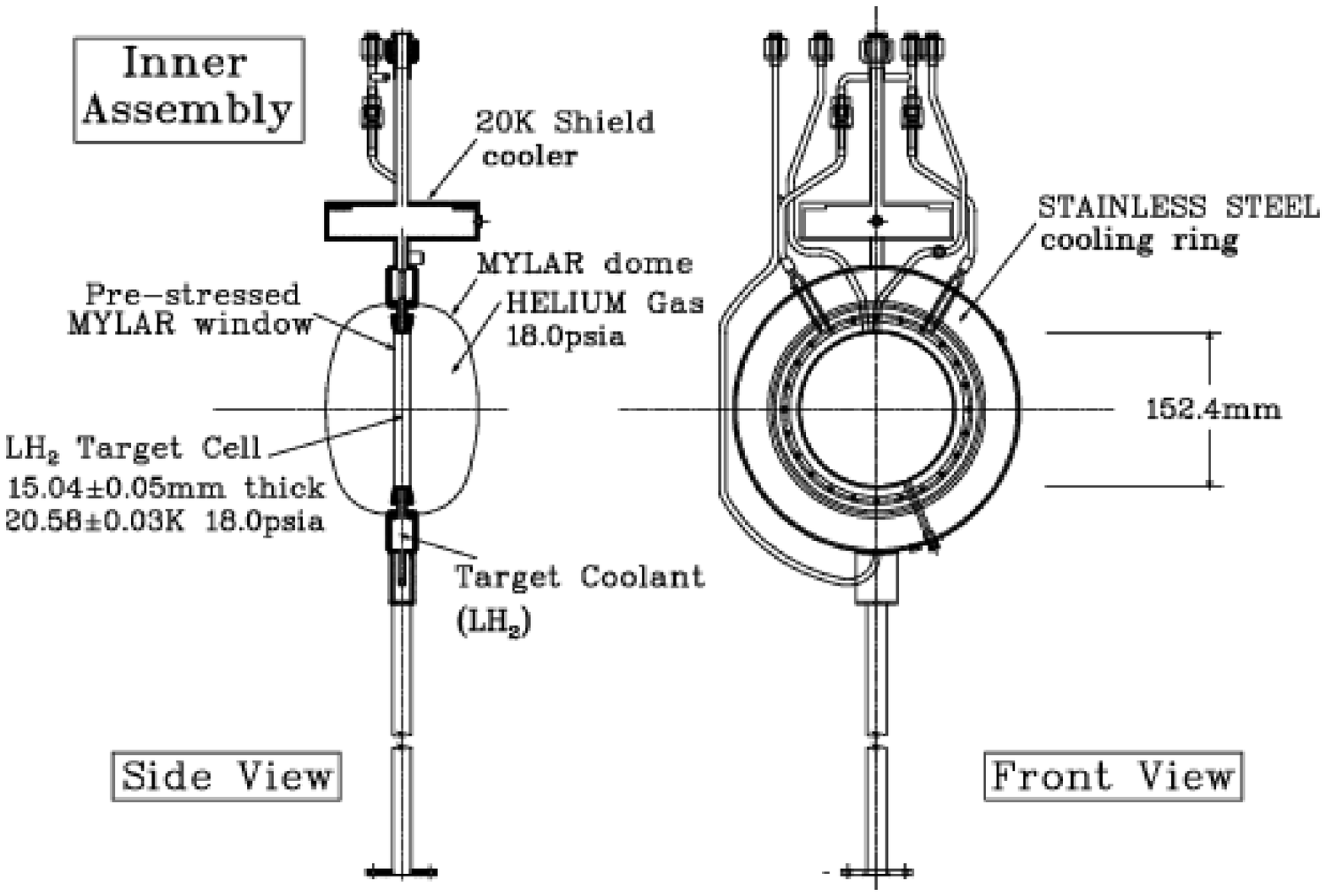,width=8.4cm} &
  \epsfig{file=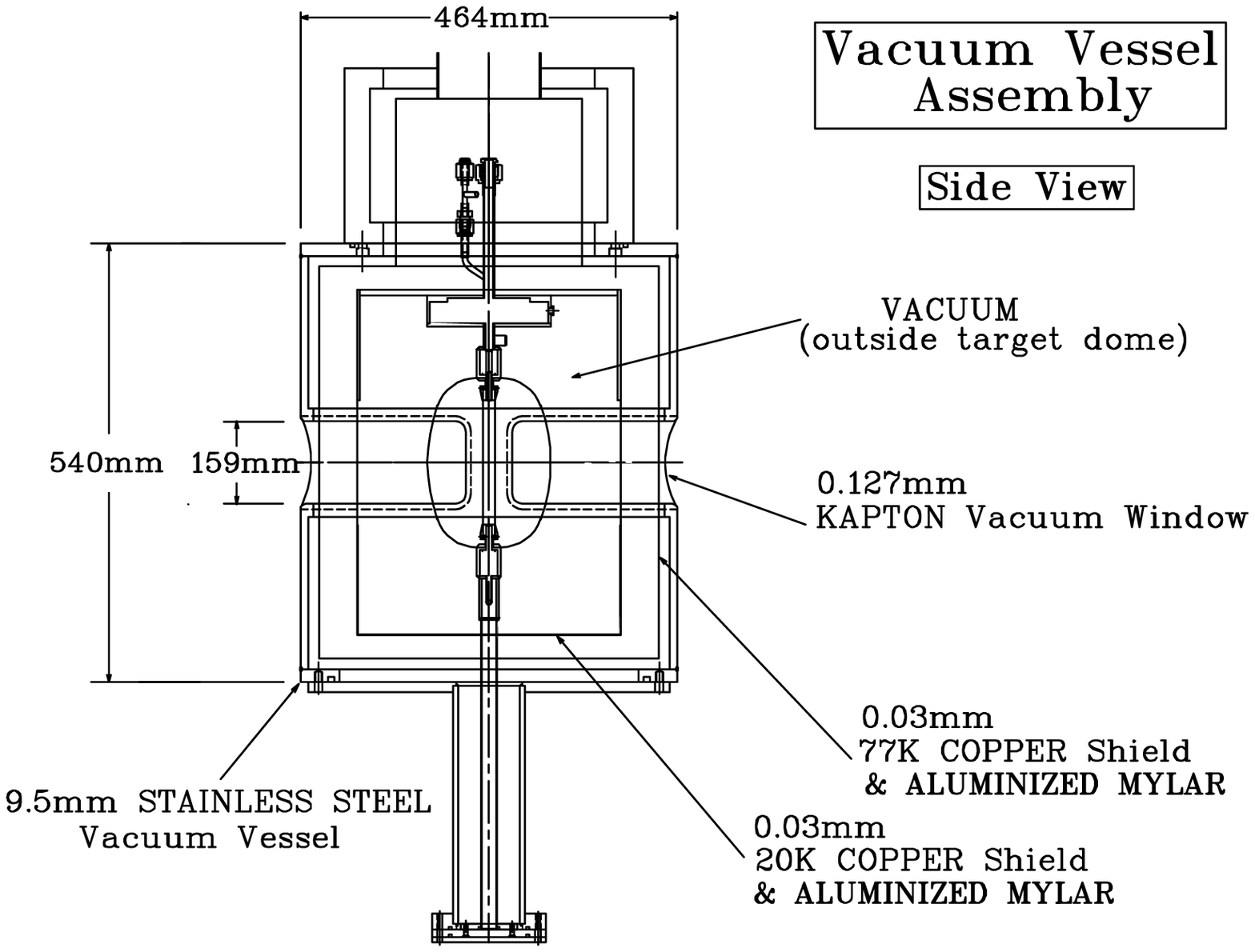,width=8.4cm} 
  \end{tabular}
\caption{
  {\bf Left}:  LH$_{2}$\ target inner flask assembly,
  showing target dimensions and operating temperature and pressure.  
  {\bf Right}: Target assembly showing
   placement of target flask assembly within outer vacuum vessel.
\normalsize}
\label{fig:lh2drawing1}
\end{figure}

A key element in the experiment was the development of the thin,
flat--walled, super-cooled liquid hydrogen target.  This target was thick
enough to provide protons at high density within a cell of accurately known
thickness, yet thin enough that energy and interaction losses to the
incoming and scattered beams were minimal, as were the corrections to the
effective solid angles due to extended source size effects.  Some
construction details of the target, including relevant physical parameters,
are displayed in Fig.~\ref{fig:lh2drawing1}.
The LH$_{2}$\/ target was contained within the 14.99$\pm$0.03 mm thick
hollow stainless steel ring and two prestressed mylar windows. The liquid
hydrogen in the target was cooled by a separate source of liquid hydrogen
flowing inside the hollow stainless steel ring.  This cooling hydrogen was
liquefied once at the beginning of the experiment and then maintained at
15.6 to 16.0 psia. The liquid hydrogen in the target itself, however, was
maintained at 18.05$\pm$0.05 psia, i.e., approx. 2.2 psia overpressure
(i.e. ``super-cooled'') in order to prevent boiling and bubbling in the
target.  The entire target assembly was contained within a large
cylindrical stainless steel vacuum vessel.  An inner copper heat shield at
the target hydrogen temperature and an outer shield at liquid nitrogen
temperature, both surrounded by aluminized--mylar superinsulation,
prevented transmission of infrared radiation onto the target and thus
further ensured that no bubbles formed. Two gaps in the vacuum vessel
covered by kapton windows provided beam access and egress.
 
Prestressed mylar windows on the target cell were used to keep the linear
thickness of the target as uniform as possible.  The deflection due to
differential pressures across the window was measured on a test bench at
liquid nitrogen temperatures as 1.83 mm/psid \cite{lh2-kelner}.  Although
quite small, this would still cause unacceptable bulging if the target cell
were contained in vacuum.  Consequently, the cell was capped on both sides
by 0.229 mm thick mylar domes containing gaseous helium at a pressure
regulated to within 10 mpsid of the pressure in the cell, causing a maximum
$\pm$0.0356 mm fluctuation in the cell width. A 140$\pm$10 mm liquid
hydrogen column above the target centre to the pressure regulation point
produced a 14$\pm$1 mpsid hydrostatic head resulting in a net
0.026$\pm$0.002 mm outward window deflection at the bottom of the target.
The helium pressure and target--helium pressure differential were digitized
and read--out online at regular intervals by the data acquisition system.
The linear thickness of the LH$_{2}$\/ target between the inside surfaces
of the mylar windows was 15.04$\pm$0.06 mm, comprised of 14.99 mm from the
machined depth of the ring, a correction taking into account shrinkage when
cooled to 20K, bulging of the windows due to the hydrostatic head, and also
a thin layer of epoxy bonding the windows to the ring. The total
uncertainty is the sum in quadrature of the individual uncertainties.
 
Target empty data for background measurements were obtained by evacuating
the target cell of all the LH$_{2}$\/ and residual gas, and replenishing it
with helium from the domes.  The helium pressure was adjusted to maintain
the same areal thickness as in target full operation: 15.8 psia at a target
angle of 53.6$^0$ and 16.1 psia at -39.4$^0$.

Due to a failure of the target vapour bulb transducer at the beginning of
the experiment, the target cell temperature could not be monitored
continuously, but was determined instead at four occasions spanning the
entire experiment by using the target cell itself as a vapour bulb.  The
vapour pressure at the LH$_2$ boiling point when the target was half full
was provided by the helium pressure transducer together with the
differential pressure transducer, both of which were regulating throughout
this process.  The resulting temperatures, inferred from vapour
pressure tables \cite{lh2-vptables}, were 20.63, 20.58, 20.56, and
20.55$\pm$0.02K, respectively, where the last value includes a small
correction which reflects the roughly 8\% ortho-(normal-) to para--hydrogen
conversion which occurred during the 16 hours after the target was filled.
The observed temperature drop is consistent with normal- to para--hydrogen
conversion in the cooling condenser fluid, which was kept at a constant
average pressure throughout the run.  The target densities at each
temperature were inferred from molar volume vs.\ temperature tables
\cite{lh2-vptables}. The average of the normal- and para--hydrogen
densities was used since the exact value of the normal/para ratio was
unknown. Although for most runs the conversion from normal to para (about
0.5\%/hour for the first 100 hours) would not have proceeded very far under
normal conditions, unknown catalytic effects might have sped up the
process.  This introduces a 0.2\% uncertainty to the target density, a
value which completely dominates that arising from the temperature
uncertainty of 0.02\% (0.01K).  Combining the measured linear thickness
together with the known average target density throughout the run, the
target areal density was determined to be 106.2$\pm$0.5 mg/cm$^2$, or
63.43$\pm$0.32 10$^{-6}$ mb$^{-1}$.

The 0$^{\circ}$ angular orientation of the target was set during the
experiment by using a transit to view markers which were placed onto the
lower rim of the vacuum vessel during target assembly.  An overall target
angle uncertainty of $\pm$0.3$^0$ was estimated based on these mechanical
measurements.  The target angle was set by rotating the entire cryostat,
with the angles read off a large disk on the support structure to an
estimated reproducibility uncertainty of 0.2$^{\circ}$.  The LH$_2$ target
angle was fixed at 53.6$^{\circ}$ for the ``Set A'' pion angle settings,
and 50.6$^{\circ}$ for the ``Set B'' settings.

\subsubsection{Tests of the LH$_2$\/ Target Angle/Thickness} 
\label{sec:angletest}

\begin{figure}
\begin{center}
  \mbox{\epsfig{file=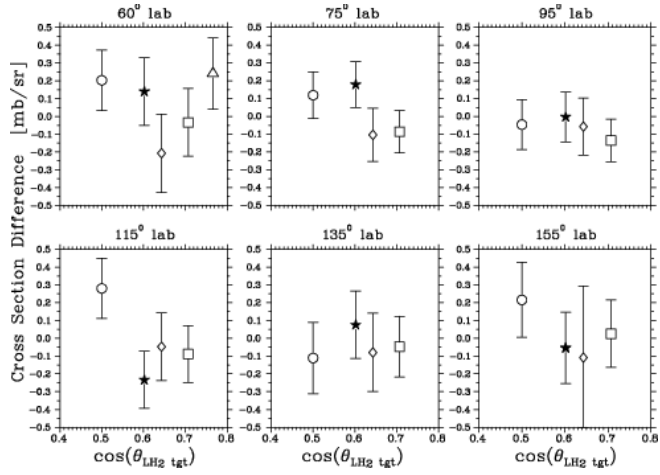, width=8.6cm}}
\end{center}
\caption{
  Two--arm $\pi^{+}$p cross sections at 169 MeV taken with the LH$_{2}$\ 
  target at angles of 45.6$^{\circ}$, 50.6$^{\circ}$, 53.6$^{\circ}$, and
  60.6$^{\circ}$. For $\theta_{\pi}$=60$^{\circ}$ lab, there is an
  additional point at -39.4$^{\circ}$.  Each graph represents one pion
  detection arm.  The ($\approx$1.3\%) uncertainties shown are statistical
  only. A target offset of +0.6$^{\circ}\pm$0.4$^{\circ}$ was inferred from
  a combination of these results together with an independent target
  thickness measurement (Fig.\protect\ref{fig:lh2_thick}).}
\label{fig:angletest}
\end{figure}

To check whether there was a systematic offset in our nominal angles,
two--arm \mbox{$\pi^{+}$p}\/ coincidence data were taken at 168.8 MeV with
nominal LH$_2$ target angles of 45$^0$, 53$^0$ (``normal''), and 60$^0$.
The ``Set B'' $\pi^+$ data (at 50$^{\circ}$) at this energy were also
considered by interpolating the data to the ``Set A'' angles. Small
uncertainties from the interpolation were added to the interpolated data.
The effect of a 0.2$^{\circ}$ target angle reproducibility uncertainty was
added to all points. The results are shown in Fig.~\ref{fig:angletest},
which include one point at $\theta^{\text{lab}}_{\pi}$=60$^{\circ}$ from
the forward angle single--arm data using a nominally -40$^{\circ}$ target
angle.  The results for each of the six pion angles were then fitted to a
form $\cos(\theta_{\text{tgt}} + \theta_0)/ \cos(\theta_{\text{tgt}})$. The
fit yielded a common offset of 0.86$\pm$0.36$^{\circ}$. Neglecting the
outlier point at $\theta^{\text{lab}}_{\pi}=115^{\circ}$ for
$\theta_{\text{tgt}}=60^{\circ}$, the offset became 0.6$\pm$0.4$^{\circ}$.

\begin{figure}
\begin{center}
  \mbox{\epsfig{file=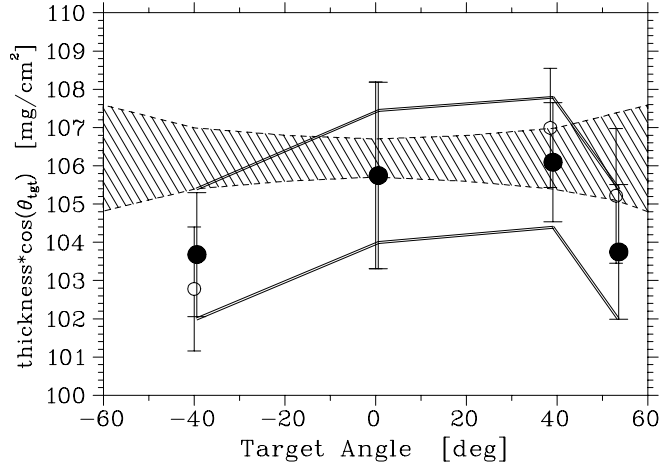,width=8.6cm}}
\end{center}
\caption{
  Results from the two different LH$_{2}$\ target thickness measurement
  methods. The solid (open) data points are the proton energy loss results
  including (ignoring) the effect of a 0.6$^0$ angular offset.  The
  $\pm$1.6\% normalization error band does {\em not}\/ include the
  contribution from the statistical errors. The hatched area represents the
  quoted target thickness uncertainty from our vapour bulb result
  ($t_{tgt}$= 106.2$\pm$0.5, Sec.~\protect\ref{sec:lh2target}) as a
  function of angle, combining the 0.5\% thickness uncertainty with the
  $\pm$0.4$^{0}$ zero--offset angle uncertainty.}
\label{fig:lh2_thick}
\end{figure}

Midway through the experiment, an independent measurement of the target
thickness was performed, involving the use of silicon counters
\cite{pnpiSi} to measure the energy loss of beam protons passing through
the LH$_{2}$\/ target. The technique is described in detail in
Ref.~\cite{mythesis}. The target was rotated to four nominal settings:
-40.0, 0.0, 38.5, and 53.0$^0$, this last angle being the setting for most
of the two-arm production runs. Fitting the data to the expression
$\mbox{X}=\mbox{X}_{0}/\cos(\theta+\theta_{0})$ yielded : X$_0$=
104.4$\pm$0.8 (stat.)$\pm$1.7 (norm.) mg/cm$^2$, $\theta_{0}$=
0.7$^{0}\pm$0.4$^0$.  Although the results from the latter three settings
were perfectly consistent with that of the vapour bulb technique
(Fig.~\ref{fig:lh2_thick}), the -40.0$^{\circ}$ point was substantially
lower, implying a systematic overall 0.7$\pm$0.4$^{\circ}$ angular
offset\footnote{The data could also be explained in terms of a +1.5$^0$
  shift at the -40$^{\circ}$ setting, since during the thickness
  measurement, the target was positioned to that angle with some
  difficulty.}. Additional evidence for a systematic angle offset was
provided by the single--arm results, which overlap better with the two--arm
results at their respective energies if an offset of about
0.5-0.7$^{\circ}$ is assumed.  Final compromise values of
$\theta_0$=0.6$\pm$0.4$^{\circ}$ with $\mbox{X}_{0}$ = 106.2$\pm$0.5
mg/cm$^2$ (from the vapour bulb result) were adopted and applied to all the
data taken with the LH$2$ target.  This value is consistent with all the
available evidence, while discounting to some extent the effect of the
outlier points at $\theta^{\text{lab}}_{\pi}=115^{\circ}$ for
$\theta_{\text{tgt}}=60^{\circ}$ in Fig.~\ref{fig:angletest}, and the
-40$^{\circ}$ point in the target thickness measurement. The resulting
uncertainty in the LH$_2$ target angle is the dominant source of
normalization uncertainty in all data taken with that target (as indicated
in Table~\ref{tab:typerrs}).

\subsubsection{Foreground and Background Running} 

Several hours were required to fill or empty the liquid hydrogen target, so
it was not possible to conduct a target empty run immediately after
completion of each target full run, or vice versa.  Therefore, a series of
target full runs was carried out for each configuration of the TOF
spectrometer and target, followed by all the respective target empty runs.
During target emptying (filling), the target would be moved out of the
beam, and the time used to conduct measurements with the solid targets.

For the case of the CH$_{2}$ targets, data runs were followed immediately
by the graphite background runs, except during those $\pi^{+}$\/ runs at
169 MeV designed to explore systematic effects. As only the two--arm
coincidence configuration was employed for the solid targets, the effect of
a relative foreground/background normalization uncertainty on the cross
sections was negligible ($<$0.1\%) due to the very low level of background
characterizing this arrangement.

\subsection{Data Acquisition}
\label{sec:dataacq}

The pulse height and timing signals from every scintillator in the system
were recorded using CAMAC electronics and read--out by computer to 8 mm
video tape using the TRIUMF\ VDACS \cite{vdacs} data acquisition program.
Both the individual and the meantime signals from the proton counters were
time digitized and scaled.  As well, all the various counter coincidences
were counted by scalers.  In particular, the BEAM output was fed into two
independent scalers as a consistency check.  The scalers accumulated
continuously when the data acquisition was active, and were read--out by
the CAMAC system at approximately one minute intervals during a run as well
as the end of a run.
 
The EVENT gate consisted of the logical OR of all six pion--proton pair
coincidences (or just pion arms for single-arm runs) together with the beam
sample signal (SAMPLE) : $\mbox{EVENT} \equiv \sum_{i=1,6} \mbox{ARM}_{i} +
\mbox{SAMPLE}$. The LAM signal, which formed the ADC gates, TDC starts, and
triggered the event readout, was the EVENT signal gated by additional
``inhibits'' depending on whether or not the computer was busy
($\overline{\mbox{BUSY}}$), whether the beam was turned off (detected using
a rate meter), or whether another EVENT signal had immediately preceded the
current one (detected using a fast inhibit).  The live time (or duty
factor) (f$_{\mbox{LT}}$) of the data acquisition system was determined
from the ratio of LAMs/EVENTs.

The ADC gate widths were set at approximately 30 ns, wide
enough to include essentially all the signal, but smaller than the beam
repetition period of 43 ns to avoid the possibility of pile--up and random
coincidences.  The TDCs operated in common start mode, with the LAM as the
common START.

In order to reduce the number of ADC and TDC channels required to
accumulate all the data from the six pion and proton arms, a multiplexing
scheme was employed, whereby the only ADC and TDC words (from $\pi$1,
$\pi$2, and P1) recorded by CAMAC were those for the arm which detected the
\mbox{$\pi$p}\/ EVENT\@.  To determine which arm caused the EVENT, the
ARM$_{i}$ timing signals for {\em each}\/ of the six arms, as well as the
beam SAMPLE signal, were fed to separate channels of an input register and
processed by the CAMAC J11 Starburst controller. This system also indicated
whether more than one arm recorded a hit for the same EVENT, thus giving
another measure of the rate of accidental coincidences. In practice, the
largest number of multiple events observed for any run was two out of many
thousand events.

 
\section{Data Analysis} 
\label{sec:data_anal}

The centre-of-mass differential cross section at laboratory kinetic energy
T$_{\pi}$ at centre--of--momentum scattering angle $\theta_{\text{cm}}$
was determined using:
\begin{equation}
 \frac{\text{d}\sigma}{\text{d}\Omega} (\text{T}_{\pi},\theta_{\text{cm}}) =
  \frac{\text{Y}(\text{T}_{\pi},\theta_{\text{lab}})\cdot
  \cos\theta_{\text{tgt}}\cdot\text{J}(\text{T}_{\pi},\theta_{\text{lab}})}{
  \mbox{N}_{\pi}\cdot\Delta\Omega_{\text{eff}}(\text{T}_{\pi},\theta_{\text{lab}}) 
  \cdot \text{N}_{\text{prot}} \cdot \epsilon }
\label{eqn:dsig}
\end{equation}
where $\text{Y}$ = number of detected \mbox{$\pi$p}\/ events at laboratory
angle $\theta_{\text{lab}}$, $\theta_{\text{tgt}}$ = target angle,
$\text{N}_{\pi}$= number of beam pions incident on target,
$\Delta\Omega_{\text{eff}}$ = effective laboratory solid angle for
\mbox{$\pi$p}\ detection, N$_{\text{prot}}$ = number of target
protons/cm$^{2}$, $\epsilon$ = scintillator efficiencies,
and J is the Jacobian transformation from the laboratory to centre-of-mass
reference frame. The target proton densities are listed in
Table~\ref{tab:target_thick}. Each of the other terms in Eq.~\ref{eqn:dsig}
are discussed separately in the following sections.  Details of the Monte
Carlo determination of $\Delta\Omega_{\text{eff}}$ are presented in
Appendix~\ref{appdx:simulDetails}, while the techniques employed for
analysis of the scintillator signals are presented in this section.  The
final cross section results are presented in Sec.~\ref{sec:results}.

\subsection{Solid Angle} 
\label{sec:geantsldang}
    
The effective solid angle of a pion arm (for single--arm operation), or
pion and proton arm combination (for two--arm coincidence mode) was
determined by Monte Carlo simulations. As the time-of-flight difference
spectra were unable to distinguish between scattered pions and those muons
arising from the decay of scattered pions, the net \mbox{$\pi$p}\/ yield
consisted of those events in the pion arm involving a pion {\em or}\/ a
muon, and a proton in the proton arm in the case of two--arm runs, all of
which needed to be modelled as faithfully as possible.  The consistency of
the simulation results with the many experimental checks that were carried
out was an important check of the procedure used and provided a useful
measure of the magnitude of many of the systematic errors characterizing
the experiment.
    
\begin{figure}
\begin{tabular}{l@{\hspace{4mm}}r}
  \epsfig{file=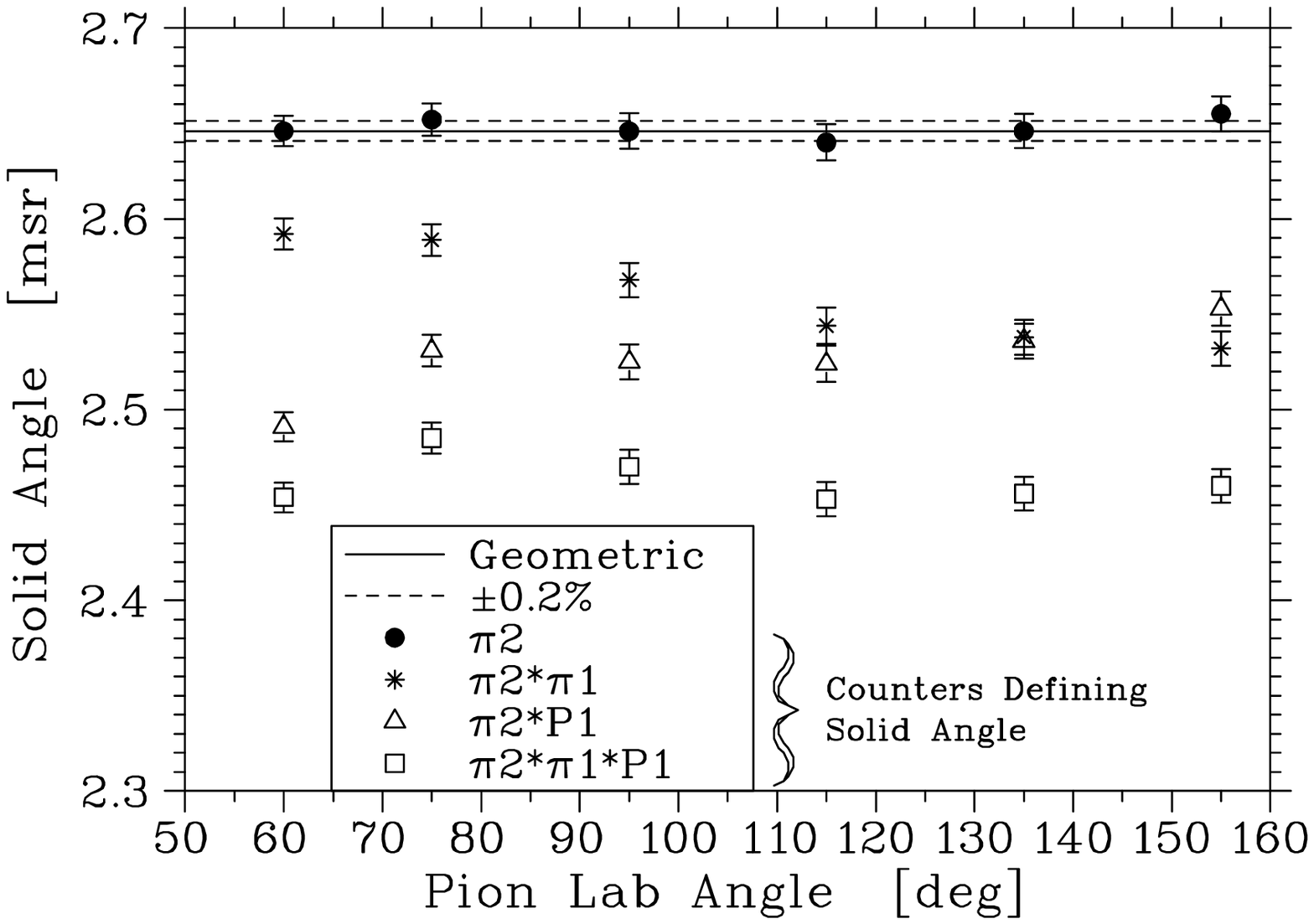, width=8.4cm} &
    \epsfig{file=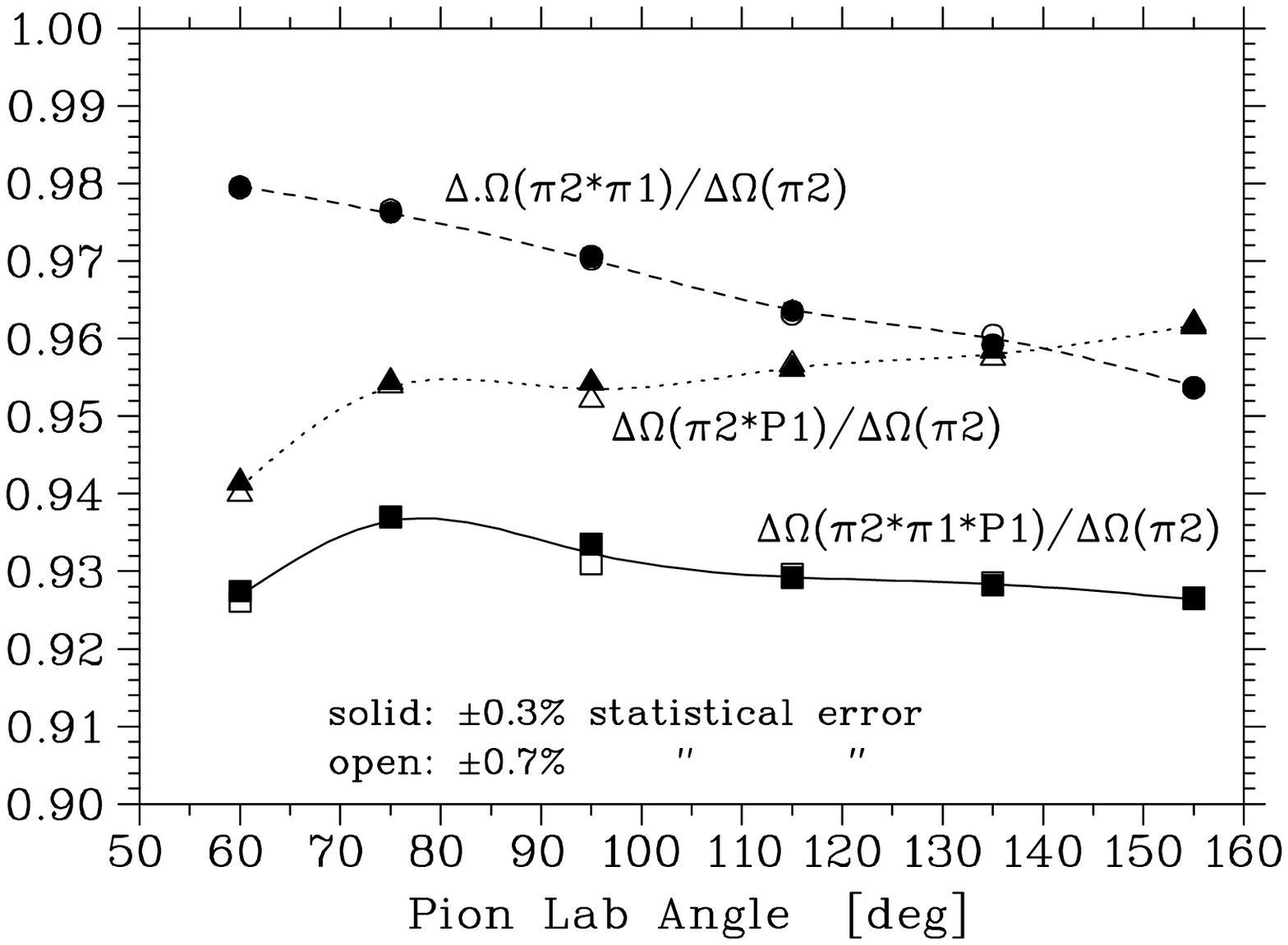,width=8.4cm} \\
\end{tabular}
\caption{
  {\bf Left:} Results of a GEANT solid angle simulation for a 141 MeV
  $\pi^{+}$p run using a 2mm CH$_{2}$\ target oriented at 53$^0$.  No hadronic
  interaction losses are included at this stage.  Note that the $\pi$2
  solid angle equals the geometric solid angle as expected.  {\bf Right:}
  Ratios of the simulated solid angles to the simulated $\pi$2 solid angle,
  showing the equivalence of the high statistics simulation with the lower
  one used in the analysis (see text).  Note that inclusion of the $\pi$1
  and $P$1 counters reduced the solid angle by only $\approx$7-8\% from
  the geometric value. The lines merely serve to guide the eye.}
\label{fig:sldang_ch2}
\end{figure}

In both the two-arm and single-arm operational modes, the solid angle
subtended by the $\pi$2 counter (2.646$\pm$0.013 msr) defined the geometric
solid angle ($\Delta\Omega_{\text{geom}}$) for detecting scattered pions.
However, some of the scattered pions (protons) that should have struck a
$\pi$2 (P1) counter failed to do so, whereas some outside the geometric
solid angle were actually detected. Consequently, an effective solid angle
$\Delta\Omega_{\text{eff}}$\/ was introduced to compensate for these
competing effects. Various factors contributed to the effective solid
angle.  A pion (proton) could suffer {\em interaction loss}\/ by hadronic
elastic or inelastic scattering on the way to the $\pi$2 (P1) counter, and
thus escape detection.  The resultant decrease in the effective solid angle
was substantial, ranging from about 2\% to 5\% depending on target and pion
scattering angles. The non-zero pion beam size resulted in an {\em extended
  source}\/ in the target from which a scattered pion could originate,
making the distance to the $\pi$2 counter (hence the solid angle) different
for each pion.  Although the effect was small, $<$0.2\%, nonetheless the
determination of $\Delta\Omega_{\text{eff}}$\/ involved the weighted
average over the extended source.  A scattered particle that would
otherwise have missed the $\pi$2 (P1) counter could suffer (Coulombic and
hadronic) {\em rescattering from an experimental structure}\/ and
subsequently hit a counter. This effect was significant only when using the
LH$_{2}$\/ target in the single pion arm setups or for the most forward
pion arm in the two-arm setup. In the former case, pions could rescatter
hadronically from the stainless steel target vessel and subsequently cause
a pion arm coincidence.  In the latter case, pions that would have missed
the pion arm counter could have scattered off the target ring and
subsequently hit the counter. In the worst cases, rescattering caused up to
$\sim$2\% effect in the effective solid angle. {\em Pion decay}\/ was
another source of pion loss, amounting to a net reduction of
$\Delta\Omega_{\text{eff}}$\/ by 2--4\% (see Fig.~\ref{fig:sldang_ch2}).
The presence of the intermediate $\pi$1 counters constrained the number of
daughter muons detected from pions that would not have been detected
otherwise. In the absence of multiple scattering, the TOF spectrometer was
designed so that for a monochromatic, point scattering source, every proton
conjugate to a pion that hit the $\pi$2 counter would hit the corresponding
P1 counter.  However, the combination of an extended source, beam momentum
spread, multiple scattering, and pion decay, spread out the recoil proton
distribution. The net result of the {\em proton counter constraint}\/ was a
(typically) 7\% reduction in the effective solid angle as illustrated in
Fig.~\ref{fig:sldang_ch2}.  Finally, particles entering near the edge of
the scintillator could exit out the side, thus reducing the path length and
consequently the light output in the scintillator. However, as most events
of this kind would have still yielded a detectable signal due to the low
thresholds used in this experiment, such corrections were expected to be
$<$0.1\%.

\subsubsection{GEANT Monte Carlo Determination of Solid Angles}
\label{sec:geanttofspect}

Due to the correlations among the various effects described above, the only
way $\Delta\Omega_{\text{eff}}$\/ could be determined accurately was by a
{\em full} Monte Carlo simulation of the \mbox{$\pi$p}\/ scattering process in
the TOF spectrometer. All the relevant physical processes associated with
\mbox{$\pi$p}\/ scattering (multiple scattering, pion decay, etc.) together
with the details of the experimental configuration (scintillators, target,
air, etc.) were included. By generating a number $\text{N}_{\text{mc}}$ of
\mbox{$\pi$p}\/ scattering events, where the scattered pion was randomly
and uniformly distributed within a solid angle $\Delta\Omega_{\text{mc}}$
chosen to be large enough to accommodate all events which could possibly
result in a $\pi$2 hit, the {\em effective}\/ solid angle
$\Delta\Omega_{\text{eff}}$\ was given by : $\Delta\Omega_{\text{eff}} =
\frac{\text{N}_{H}}{\text{N}_{\text{mc}}} \cdot \Delta\Omega_{\text{mc}}
\quad \bigl(\pm \frac{100}{\sqrt{\text{N}_{H}}} \% \bigl)$ where
$\text{N}_{H}$ was the number of $\pi$2 counter hits in the simulation. The
uncertainty is that expected from the Poisson limit to the applicable
binomial statistics.

The GEANT \cite{geant} detector description and Monte Carlo particle
tracking program was used to simulate the \mbox{$\pi$p}\/ scattering
reaction in the TOF spectrometer for every {target, target angle, pion
  angle, trigger} configuration.  All elements of the TOF spectrometer were
accurately modelled, including composition, dimensions, and positions of
all in--beam, pion arm, and proton arm counters as well as relevant
characteristics of the CH$_2$ targets (when employed). The LH$_{2}$\/
target was also faithfully modelled, including all windows, dome, heat
shield and super--insulation layers, stainless steel vacuum vessel, target
cooling ring, and liquid hydrogen coolant in the ring (see
Fig.~\ref{fig:lh2drawing1} and others in \cite{mythesis}).  Accurate
modeling of this kind ensured proper treatment of the effects due to
interfering structures that were discussed previously. All the relevant
physical processes were included in the simulation, though in most cases
the hadronic interaction routine was not, since it was found to be too
imprecise for calculating nuclear absorption losses in the low energy
region relevant for this experiment.  These small interaction losses were
subsequently introduced by hand to obtain the final results. Details of the
simulation are described in Ref.~\cite{mythesis}, while some details
concerning the hadronic interaction corrections introduced subsequent to
the discussion found in Ref.~\cite{mythesis} are presented in
Appendix~\ref{appdx:simulDetails}.

\subsubsection{Tests of the Effective Solid Angle Determination} 
\label{sec:ch2thicktest}
\label{sec:sing-vs-coinc}

The solid angle results from the GEANT simulations showed that effects of
the $\pi$1 and/or P1 counters affected the geometric solid angle subtended
by the $\pi$2 counter by no more than $\approx$9\% in the {\em worst}\/
case, and more typically by only $\approx$5\%, while the other effects
(multiple scattering, etc.) were even smaller. Thus {\em a priori}\/ it
would be expected that the systematic uncertainty introduced by these
various effects should be smaller than about 1\% (i.e., 10\% uncertainty of
the corrections).

\begin{figure}
\begin{center}
\mbox{\epsfig{file=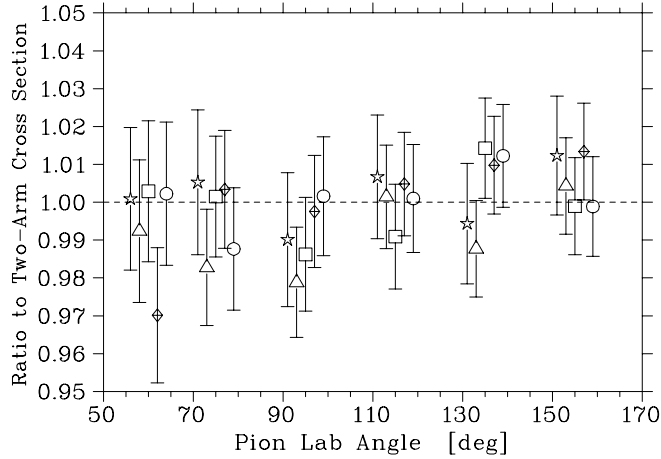, width=8.6cm}}
\end{center}
\caption{
  $\pi^{+}$p cross section ratios to the normal two-arm value
  at 169 MeV using the LH$_{2}$\ target, measured for various
  configurations at different times during the experiment. For clarity, the
  angles are offset from one another, and the uncertainties shown are
  statistical only. The boxes (crossed diamonds) refer to runs when the
  proton arm radius (angle) was shifted, the stars (circles) when the
  proton P1 (pion $\pi$1) counter was removed from the coincidence, and the
  triangles when the beam momentum spread was increased to 3\% (from 1\%).
  These results demonstrate the insensitivity of our technique to such
  systematic effects.}
\label{fig:syst_checks}
\end{figure}

Nevertheless, to test the reliability of the effective solid angle
determinations, \mbox{$\pi^{+}$p}\/ data sets at 168.8 MeV were obtained
under various experimental setups which differed from the standard two--arm
configuration. For example these included deliberately misaligning the
proton P1 counters by +0.25$^0$, changing the P1-target distance from 920
mm to 855 mm or 1020 mm, removing the $\pi$1 counter from the EVENT
coincidence, removing the P1 counter from the EVENT coincidence (i.e.\
single--arm mode), and increasing the incident beam momentum spread to 3\%
$\Delta$p/p (from 1\%). The resulting cross-sections are shown in
Fig.~\ref{fig:syst_checks}. Although the effective solid angles varied by
as much as 5\% among the various configurations, it is evident that all the
data are consistent at about the 1\% level.

\begin{figure}
\begin{center}
  \mbox{\epsfig{file=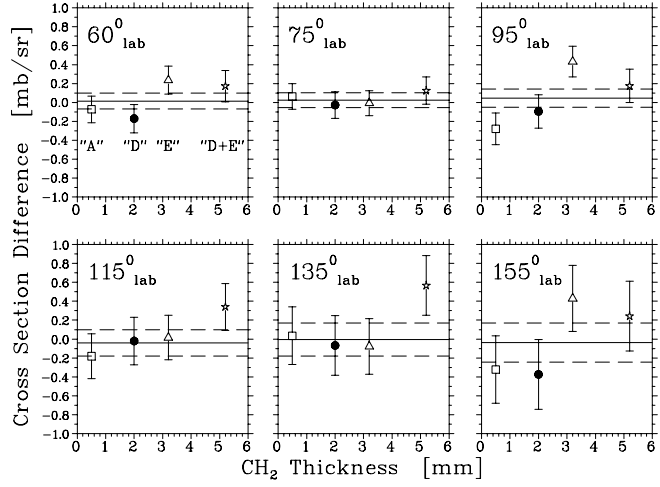, width=8.6cm}}
\end{center}
\caption{
  $\pi^{+}$p cross section differences at 169 MeV for the two--arm
  setup using CH$_{2}$\ targets of various thicknesses.  The uncertainties
  shown are purely statistical. Not shown are normalization uncertainties
  dominated by the target proton density (1\%) and the $\pm$0.2$^0$ target
  angle error (0.5\%). The horizontal solid and dashed lines represent the
  weighted average and uncertainty ($\approx$0.8\%) for the three targets
  ``A'', ``B'', ``D''.  The solid points are the data used for the final
  results at this energy. These results confirm the quoted 1.4\% relative
  density uncertainty for these targets.}
\label{fig:thicktest}
\end{figure}

In addition, two--arm \mbox{$\pi^{+}$p}\/ coincidence runs at 168.8 MeV
utilizing different solid CH$_2$ targets were obtained to provide
information concerning the relevant sizes of the uncertainties associated
with the target proton density (1\%), the effect of hadronic interaction
losses in the target on the incident beam normalization, and the effect of
varying scattered particle multiple scattering and hadronic interaction
losses on the solid angle determination.  Data were accumulated for
CH$_{2}$\/ target thicknesses of $\approx$0.5 mm (target ``A''), 2.0 mm
(``D''), 3.2 mm (``E''), and 5.2 mm (``D''+ ``E'').  The relative cross
section differences for each of the targets are shown in
Fig.~\ref{fig:thicktest}.  The error bars shown are purely statistical. The
beam pion losses varied from 1.2\% for target A to 2.4\% for target E+D,
while the modification of the effective solid angles associated with these
targets varied between 6.5\% and 1.5\% at the extremal angles, due to
different proton energy losses and hadronic interaction losses.  The
results are completely consistent within the relative normalization
uncertainties. The solid and dashed horizontal lines in
Fig.~\ref{fig:thicktest} are the weighted averages of the cross section
using the three thinnest targets and are included to better visualize the
results.  Considering the order of magnitude change in target thickness
between target A and E+D, these results provide confidence that the solid
angles and other target-dependent uncertainties are well understood.

\subsection{Beam Intensity Determination} 
\label{sec:beamnorm}

The number of incident pions N$_{\pi}$ per run involved a product of six terms:
\begin{equation}
\text{N}_{\pi} = \text{B}\cdot \text{f}_{\text{LT}}\cdot \text{f}_{\pi}\cdot
                  \text{f}_{\text{D}}\cdot \text{f}_{\text{L}}\cdot
                  \text{f}_{\text{S}}
\label{eqn:dsigbeam}
\end{equation}
where B = BEAM coincidences recorded by hardware scalers, f$_{\text{LT}}$ =
data acquisition live time fraction (efficiency), f$_{\pi}$ = pion fraction
to the channel exit, f$_{D}$ = pion survival fraction (decay) from channel
exit to target centre, f$_{L}$ = pion survival fraction after interaction
losses to target centre, and f$_{S}$ = correction factor accounting for
multiple pions traversing the target. Each of these factors is discussed in
detail in the following sections.

\subsubsection{Beam, B and livetime, f$_{\text{LT}}$}

The number of BEAM coincidences, B, was counted in two separate CAMAC
scaler modules. As mentioned in Sec.~\ref{sec:dataacq}, the live time
f$_{\text{LT}}$ was measured as the ratio of the CAMAC scalers LAMs/EVENTs.
These values were checked using visual scalers. Since all three scalers
recorded large values with no discrepancies, B and f$_{\text{LT}}$ were
considered to be virtually error--free.

\subsubsection{Pion Fraction, f$_{\pi}$ }
\label{sec:beamcomp}
   
Determination of the pion fraction f$_{\pi}$ proceeded in two steps:
removal of the proton contribution in the beam, and determination of the
beam contamination due to muons and electrons originating near the pion
production target.  Protons were effectively removed from the BEAM by means
of differential energy loss within the channel provided by a midplane
absorber before the second bending magnet, and a $\overline{\text{S2BH}}$
upper level discriminator used to reject the residual (large pulse height)
protons.  However, at the highest two beam energies involved in the
experiment, some protons managed to leak into the BEAM definition.  These
protons were easily identified by the TCAP SAMPLE spectrum and large pulse
heights in all the in--beam counters.  As the corrections due to these
protons were never large ($<$0.5\%), their presence only introduced a small
additional uncertainty to f$_{\pi}$ ($<$0.1\%).

\begin{figure}
\begin{center}
  \mbox{\epsfig{file=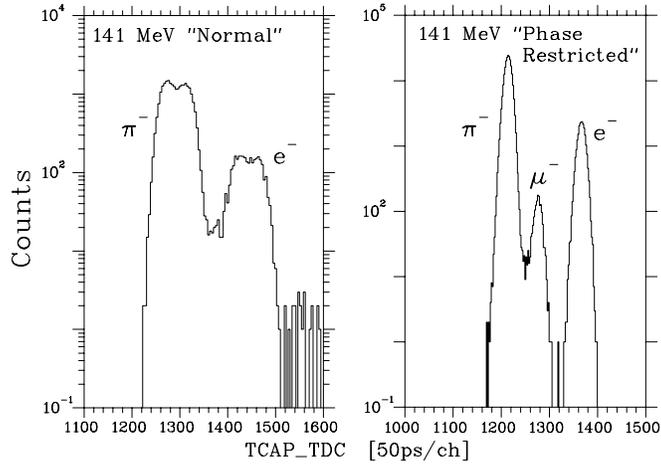,width=8.6cm}}
\end{center}
\caption{
  {\bf Left:} Normal pion beam time structure measured with respect to the
  TCAP probe as exhibited during a 141 MeV $\pi^{-}$p run.  The electrons
  are clearly separated, but the muons are totally obscured under the pion
  and electron peaks.  {\bf Right:} TCAP time spectrum, obtained during a
  dedicated run using phase--restricted primary proton beam on the pion
  production target.  In this case, the pions, muons, and electrons are all
  clearly distinguished.  }
\label{fig:norm-vs-prt}
\end{figure}

The muon and electron contamination of the pion beam arising from pion
decay near the pion production target was readily determined using particle
TOF to the S2B counter.  At low beam energies ($\leq$ 110 MeV), the length
of the channel allowed all three components to be easily resolved during
normal operation \cite{bra86}.  However at higher energies, the poor time
resolution imposed by the normal 3 ns width of the proton beam buckets
limited the clean separation of particles, and, for most of the energies
involved in the experiment, this limitation meant that the muon (and at the
highest energies even the electron) components were obscured by the
dominant pion peak. Consequently, a series of runs with a reduced proton
pulse width were dedicated to measuring the beam composition. As
illustrated in Fig.~\ref{fig:norm-vs-prt}, the $\approx$1 ns wide Gaussian
distribution achieved in this ``phase--restricted'' mode of operation was
significantly narrower than the $\approx$3 ns double--Gaussian time
structure seen during normal runs.
   
Two sets of phase-restricted runs were carried out at all the energies
involved in this experiment, using the same pion production target and
midplane absorbers employed in the data production runs.  The midplane
slits were set to 0.5\% $\frac{\Delta p}{p}$ in the first series of
measurements, and 2\% in the second, with the jaws set at a value midway
between those used in the normal $\pi^+$ and $\pi^-$ runs. In addition,
some runs were obtained with the momentum slit widths varied between 0.5\%
and 2.5\%.  Since only beam counter information was required for these
tests, all pion and proton arms were removed from the EVENT definition.
The $\pi$:$\mu$:e ratios were determined from Gaussian fits to the TCAP
timing spectra. All three peaks were cleanly separated except at the
highest two energies, where the muon peak was obscured by the tail of the
pion distribution. However, by fixing the muon peak position using the
expected $\pi$--$\mu$ and $\mu$--e TOF differences at those momenta, robust
fits were found. The electron component could be clearly separated at all
energies.
   
\begin{figure}
\begin{center}
  \mbox{\epsfig{file=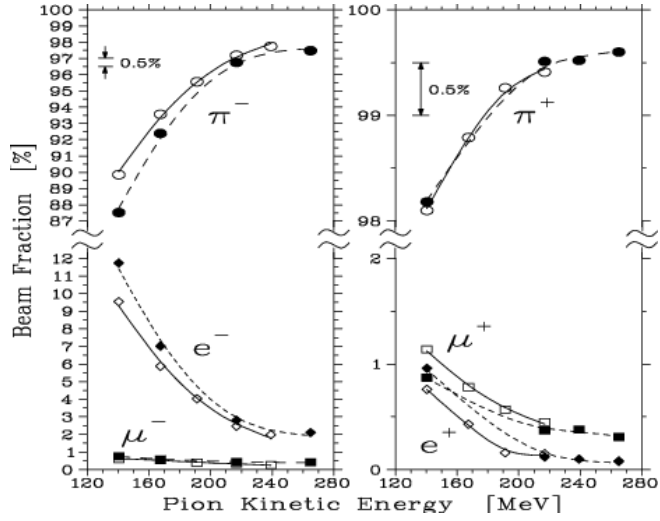, width=8.6cm}}
\end{center}
\caption{
  Percentage of pions, muons, and electrons in the M11 beam as defined by
  the in--beam scintillators during two runs (run 1 open ; run 2 filled
  points) with phase-restricted primary proton beam. The differences
  between the series are due in part to changing electron contamination
  from different midplane slit settings (see text). The relative
  $\pi$/($\pi$+$\mu$) fraction is constant ($\leq$0.2\% difference at 141
  MeV) between the two series.  The beam composition was found to be
  insensitive to typical drifts of the proton beam location on the
  production target. The lines merely serve to guide the eye.}
\label{fig:prt_fractions}
\end{figure}

The pion fraction results (f$_{\pi}$) from the two series are summarized in
Fig.~\ref{fig:prt_fractions}. Immediately obvious is the difference in the
$\pi^{-}$\/ results between the two series, ranging from 2\% at 140 MeV to
0.5\% at the highest energy.  On the other hand, the $\pi^{+}$\/ results
from the two series differed by less than 0.2\% at all energies. Such
results are not unexpected. As the muons originate from a more distributed
source at the production target, and so are not focussed as well as the
pions at the midplane, the $\pi$/($\pi + \mu$) fraction should depend on
midplane slit setting. Consequently, narrow slit settings favour the pions
over the muons.  This effect was expected to be even larger for the
electrons, since their source was even more spatially distributed, and they
were a much larger fraction of the $\pi^{-}$\/ beam than were the muons.
The experimental results confirmed this.  As there were many fewer muons
and positrons in the $\pi^{+}$\/ beam, such effects were much less
significant there.  The REVMOC \cite{revmoc} simulations also showed that
the fractions depended on the geometry of the beam counters, since larger
(smaller) counters would intercept a larger (smaller) fraction of the muons
and electrons which tend to form a halo around the pion beam.  Because of
this dependence, our results cannot be compared directly to those of
Ref.~\cite{smith-prt} where the phase--restricted beam technique was also
used. However, the trends observed here are consistent with those shown in
that work.  Tests with the position of the proton beam deliberately varied
on the production target showed that the above results were insensitive to
the typical amount of beam variation monitored by our beam hodoscope during
our runs.

\begin{figure}
\begin{center}
  \mbox{\epsfig{file=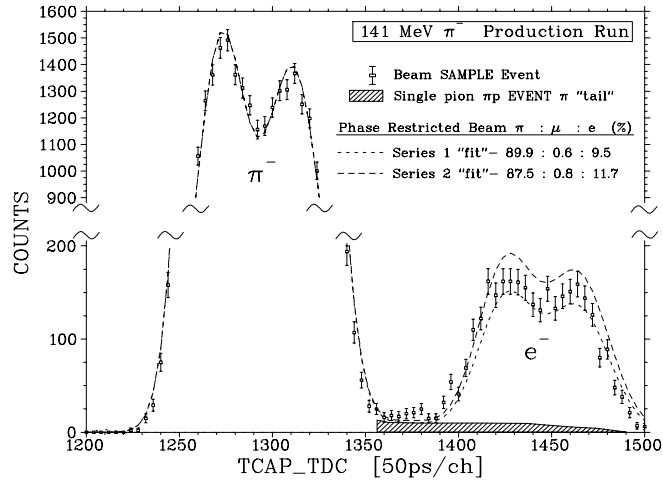, width=8.6cm}}
\end{center}
\caption{
  Example of a beam sample TCAP timing spectrum taken during normal
  production running. Two Gaussians are fit to the pion peaks, and the
  electron and muon (obscured by the $\pi$ and $e$ peaks) contributions are
  estimated using the results obtained from the phase restricted beam runs.
  This technique was used to estimate f$_{\pi}$ for the production runs.
  The tail to the pion timing distribution is inferred from true $\pi$p
  coincident events associated with only a single beam pion per bucket.}
\label{fig:tcap_normalfit}
\end{figure}

The results shown in Fig.~\ref{fig:prt_fractions} demonstrate that the
$\pi^{+}$ fraction was $\geq$98\% at all energies, and the results of the
two series were consistent to $\approx\pm$0.1\%.  These results were used
for $f_{\pi^+}$ at all energies. However, in the case of \mbox{$\pi^{-}$p},
multiple Gaussian fits were performed on the TCAP spectra during normal
operation in order to fine-tune the results of the run in question.
Although the muon contribution in the TCAP spectra was obscured by the pion
peak during normal running, the electron peaks could be easily identified
for incident pion energies up to 218 MeV.  Because the time structure
during normal operation resulted in 2 (sometimes 3) well-defined timing
peaks, multiple Gaussian line shapes were fit to the pion peaks, with the
phase-restricted beam results used to constrain the muon and electron peak
amplitudes and centroids. A typical fit overlayed on the TCAP beam SAMPLE
spectrum is illustrated in Fig.~\ref{fig:tcap_normalfit}. During normal
beam operation, the timing spectra possessed small ``right hand'' tails.
The fact that these tails were due to pions and beam contamination was
inferred from the fact that they were apparent for single $\pi$p events, of
the form $\Pi\cdot$P$\cdot\overline{\text{VETO}}$ EVENT, since only pions
could yield a (pion arm)--(proton arm) coincidence. By fitting the electron
peak together with the pion tail, the electron fraction could be readily
determined. As the phase--restricted beam results indicated that the
$\pi$/($\pi$+$\mu$) ratio varied by less than 0.2\%, f$_{\pi}$ could be
reliably estimated.  In all cases, the $\pi^{-}$\/ results were within the
range of values of the two phase-restricted beam series, whereas the
$\pi^{+}$\/ results were within 0.2\% of the phase-restricted beam values.
At the lower energies where the electron component could be reliably
estimated, the results for f$_{\pi}$ extracted in the above manner were
used in the cross section calculations, whereas at the higher energies, the
phase-restricted beam results were used.  Since the $\pi^{-}$\/ results
were always between the results for the two phase-restricted beam series,
the estimates at the higher energies were not expected to vary by more than
$\approx$0.5\% from those shown in Fig.~\ref{fig:prt_fractions}.

\begin{figure}
\begin{center}
  \mbox{\epsfig{file=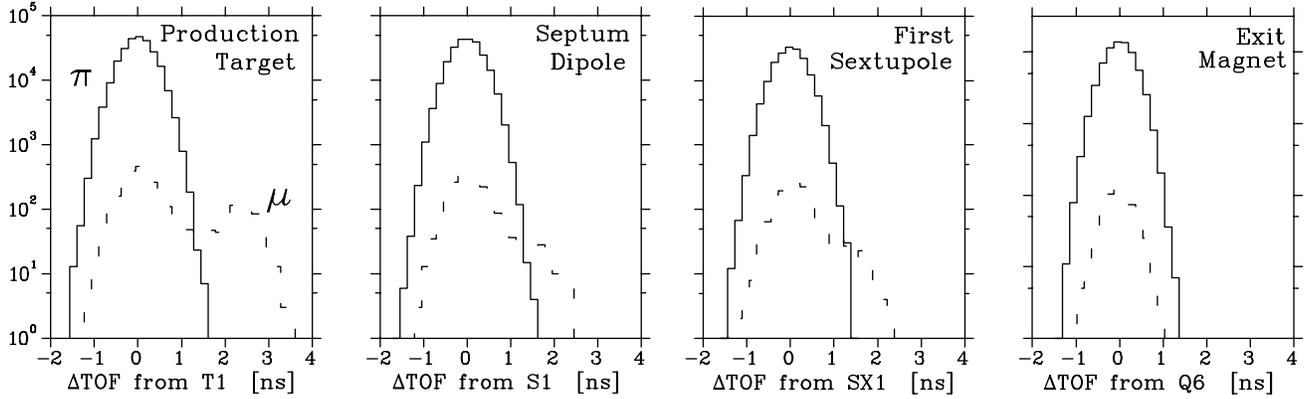,width=17.2cm}}
\end{center}
\caption{
  REVMOC simulations of 274 MeV/c $\pi$--$\mu$ time--of--flight differences
  starting at different points along the pion channel. Most of the muon
  peak at right originates near the production target, with little
  contribution ($\approx$0.2\%) between the septum (first magnet after
  production target) and the channel exit.  After the channel exit, the
  muon contamination under the pion peak is easily determined by REVMOC or
  GEANT. The number of simulated events in each spectrum is not identical.}
\label{fig:rev_m11-inchannel}
\end{figure}

The precise nature of the quantity, f$_{\pi}$, should be emphasized here.
Since muons from those pions which decay {\em after} the channel exit could
not be distinguished from pions by TOF (see
Fig.~\ref{fig:rev_m11-inchannel}), the ``pion fraction'' defined above
contains a contribution from these decay muons.  The fraction f$_{\pi}$
thus represents the fraction of the beam at the last channel element that
consists of those pions, muons, and electrons that would {\em
  subsequently}\/ cause a BEAM coincidence. However, to obtain the fraction
of BEAM coincidences due to {\em pions} in the {\em target}, f$_{\pi}$ had
to be corrected for pion decay in the channel and downstream of the channel
exit (f$_{D}$) and for hadronic interaction losses (f$_{L}$).

\subsubsection{Pion Decay, f$_{D}$} 
\label{sec:beamdecay}

Pion decay downstream of the last channel element, f$_{D}$, was calculated
in a straightforward fashion by Monte Carlo simulation using the GEANT
\cite{geant} program. Neglecting hadronic interaction losses, the factor
f$_{D}$ represented the fraction of the pions at the exit (quadrupole)
magnet midplane which would subsequently produce a BEAM coincidence and hit
the target.  Since this factor was dependent on the beam counter geometry
and beam phase space, the simulations were carried out for each run using
the beam phase-space parameters measured during the channel tuning phase of
the experiment.  The results for f$_{D}$ varied between 0.961 at 141 MeV
and 0.973 at 267 MeV, and were rather insensitive to the beam size.
Considering the decay correction only up to the S2B counter, the pion
survival fraction was 0.992, virtually independent of energy and beam size.
As the energy increases, fewer pions decay, but since the muon cone angle
is smaller, more of the decay muons are detected. These GEANT results for
the pion decay to the S2B counter and to the target were checked using the
REVMOC \cite{revmoc} beam transport program, with the two simulations
agreeing to better than 0.1\% at all energies.
 
\begin{figure}
\begin{tabular}{l@{\hspace{4mm}}r}
  \epsfig{file=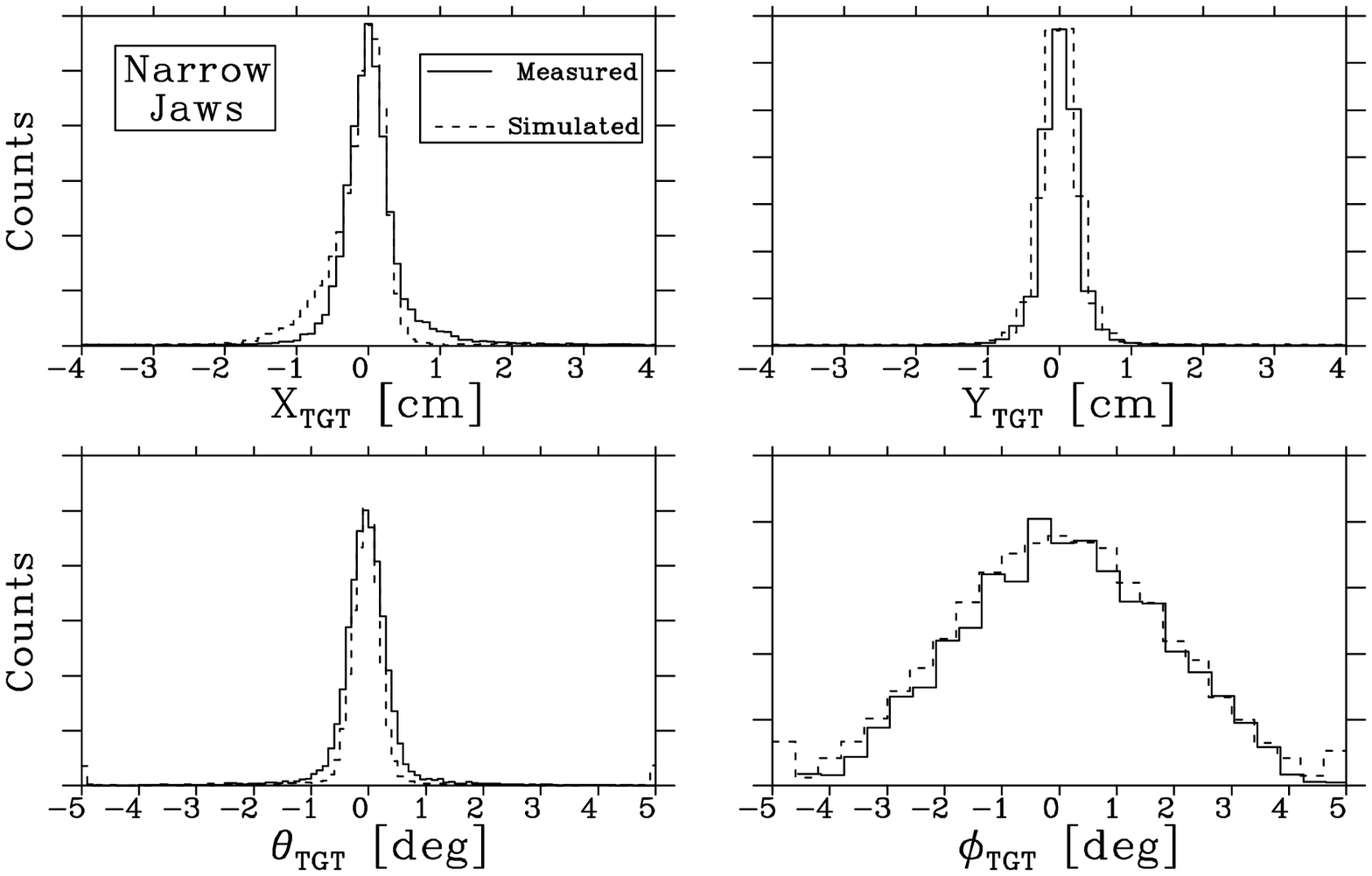,width=8.4cm} &
  \epsfig{file=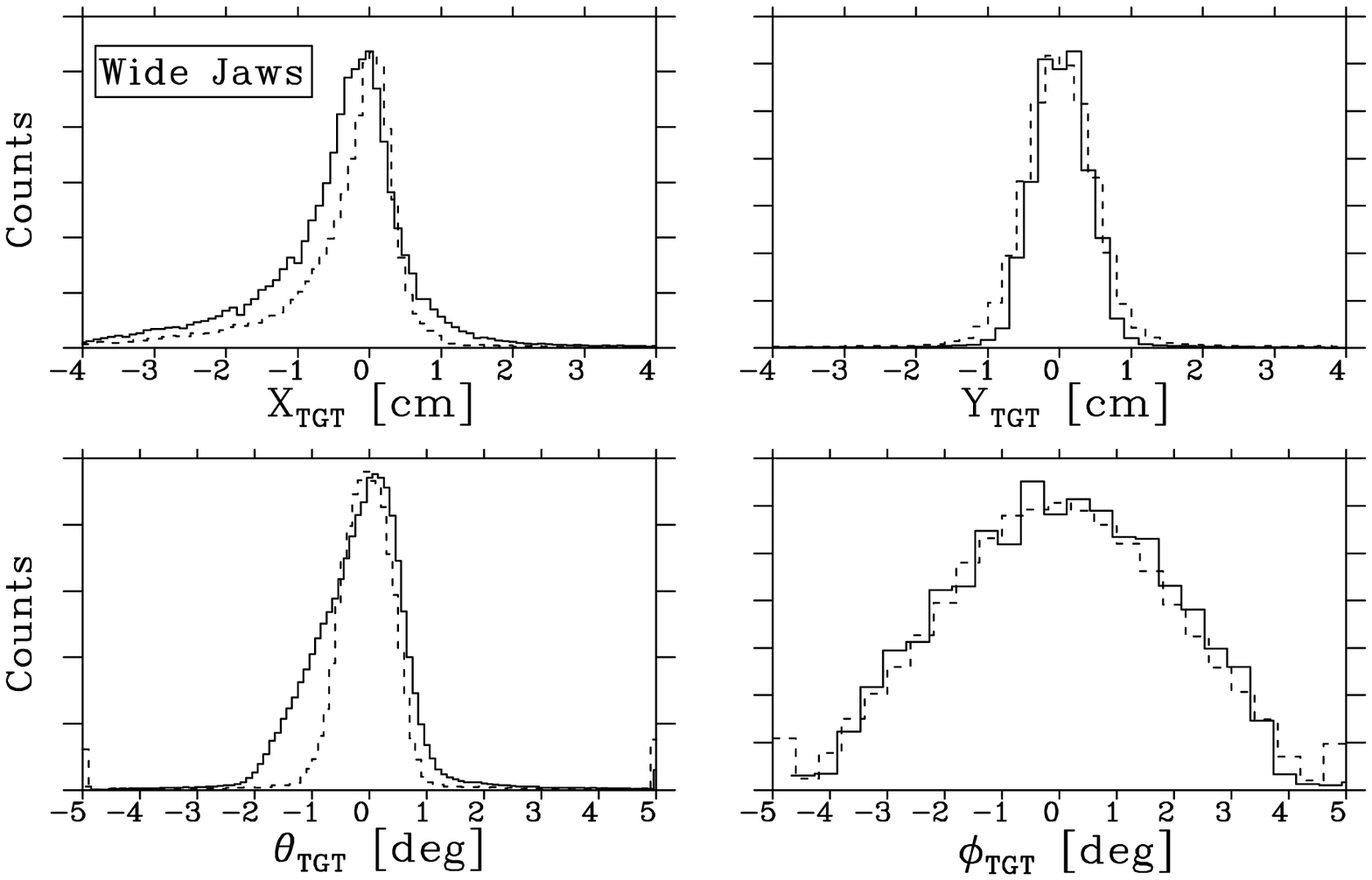,width=8.4cm} \\
\end{tabular}
\caption{
  Comparison of the incident pion beam size and divergence measured just
  prior to the experiment for two settings of the channel front--end
  rate--restricting jaws ({\em narrow}: H=50cm, V=30mm ; {\em wide}:
  H=140cm, V=120cm).  Overlayed are predictions from a REVMOC simulation of
  the channel. The satisfying agreement lends confidence to our
  beam--related corrections (e.g. in--channel pion decay,
  Fig.~\protect\ref{fig:rev_m11-inchannel}).  }
\label{fig:rev-vs-beam}
\end{figure}

One source of muons not yet discussed was that originating from pion decay
within the channel.  As the muons appearing in the muon peak possessed the
largest TOF difference with respect to the pions, they originated near the
production target.  Since the $\pi-\mu$ TOF difference for muons
originating after the last channel magnet was smaller than the instrumental
timing resolution, they would consequently appear under the pion peak.
Muons originating between the production target and the last magnet would
have a timing distribution spread between these two extremes. The results
from the lower energy phase-restricted beam fits show that the size of this
contribution was significantly smaller than that from the muons originating
from the production target, since a clear gap existed between the pion and
muon peaks.  The contribution was expected to be small since most decay
muons emerge at an angle to the direction of the incident pion (the
``Jacobi angle'', varying from $\approx$9$^0$ at 140 MeV to $\approx$6$^0$
at 270) with a momentum different from the pion, so these muons would
either strike the beam pipe walls or be bent away from the rest of the beam
by the magnets.  Nevertheless, a Monte Carlo simulation of the pions and
their decay muons from the production target through the beam line to the
in--beam counters and scattering target was undertaken using the REVMOC
\cite{revmoc} beam transport program \cite{mythesis}.
Fig.~\ref{fig:rev-vs-beam} shows the pion beam phase space parameters for
two settings of the rate restricting jaws, demonstrating that the channel
was well understood.  This simulation confirmed that the contamination from
muons originating between the production target and the channel exit was
small, amounting to only about 0.2$\pm$0.1\%. This is seen clearly in
Fig.~\ref{fig:rev_m11-inchannel}, which shows the time distribution of
pions and muons originating from various points in the channel.  Therefore
the correction f$_{\pi}\rightarrow 0.998\cdot$f$_{\pi}$ was utilized in the
beam normalization factor for all data--taking runs.

\subsubsection{Pion Beam Interaction Losses, f$_L$} 
\label{sec:beamloss}
  
Although the simulations discussed above neglected pion loss through
hadronic interaction since neither the REVMOC nor GEANT simulations of
these interactions were sufficiently reliable for this application, the
factor f$_{\pi}$ {\em already included}\/ hadronic loss effects to the S2B
counter, since it was measured experimentally in the phase-restricted beam
runs.  Therefore, the only hadronic losses of pions that remained to be
accounted for in the beam normalization factor were those in the CH$_2$ or
LH$_2$ targets (and shields, windows, etc.\ in the case of the LH$_{2}$\/
target), in the air, and in the S2B counter ({\em after}\/ a hit had been
registered by the electronics).  All the pions which interacted in the air
and in the various shields and windows surrounding the LH$_{2}$\/ target
were assumed lost, since it was very unlikely that such pions would
continue forward to the target.  For the S2B counter, pion losses in the
tape on the downstream face of the counter and in the final third of the
scintillator were included. The maximum total hadronic loss correction
calculated in this way was $\approx$2.0\% for the case of 168.8 MeV
$\pi^{+}$\/ in the LH$_{2}$\/ target oriented at 60$^0$, with the largest
contribution being that of pion loss to inelastic channels in the target.
The uncertainty in this value due to the various assumptions and hadronic
cross section estimates was estimated to be about 10\% arising from the
uncertainties in the pion--nucleus cross section estimates with an
additional 10\% due to the assumptions concerning the interaction losses,
implying a loss correction for the above case of $\sim$2$\pm$0.3\%.

\subsubsection{Multiple Pion Correction, f$_S$} 
\label{sec:multpion}

For a $\sim$1 MHz pion rate, implying delivery of a pion to the target area
in only one out of every 25 beam buckets, the problem of signal pile-up
characteristic of high intensity, low duty factor accelerators was
negligible, since at these rates, using Poisson statistics, the
probability of multiple pions in a single beam bucket was only about 3\%.

A BEAM coincidence would indicate only that {\em at least one} particle had
passed through all the beam counters, regardless of how many pions were
delivered in a single beam bucket. Therefore, when calculating cross
sections, either the total beam counts, B, had to be increased
appropriately to correct for events with more than one incident pion, or
these events had to be rejected outright from the analysis. As a
consistency check, both approaches were employed when determining the
single--pion beam correction factor, f$_{S}$.

The first of these methods, involving correcting the beam counts, B, was
called the ``Poisson'' correction scheme since the number of pions
occurring in a single beam bucket was described by Poisson statistics.
In beam counter geometries where all the pions which could have reached the
target must have passed through the counters, the correction to B would
have been straightforward: if the probability of a pion occurring within a
beam bucket is $\lambda_{B}$ (typically several percent), and $\nu$ is the
frequency of the beam buckets (23.058 MHz), then the actual rate of pions
traversing the beam-defining counters would be $\lambda_{B}\cdot\nu$,
whereas the rate indicated by the counter scalers would be the rate of one
or more pions per bucket, or $\nu\cdot(1-\text{e}^{-\lambda_{B}})$.
Therefore, the correction factor for the beam-defining counters would have
been $\text{f}^p_{S}= \lambda_{B} /(1-\text{e}^{-\lambda_{B}})$

In our case, however, a small portion of the pion beam {\em missed}\/ the
beam-defining counters, yet still traversed the target.  Consequently, if
one or more pions traversed the target {\em in addition}\/ to the one which
travelled through the beam counters, the chance of a $\pi p$ interaction
would have increased correspondingly.  If the probability per beam bucket
of finding a pion able to traverse the target was $\lambda_{T}$, (thus a
true pion rate on target of $\lambda_{T}\cdot\nu$), then the probability
that the pion {\em also}\/ passed through the beam-defining counters was
$\rho = \lambda_{B} / \lambda_{T}$ and the probability that it missed was
$\gamma = 1- \rho$. In this case, the actual pion rate traversing the
target {\em and}\/ associated with a count in the beam counters is
$\text{R}^{\pi}_{\text{beam}}$=$\nu\lambda_{T}(1- \gamma
\text{e}^{-\lambda_{B}})$, which replaces the $\lambda_{B}\cdot\nu$ used
previously. Thus the multiple pion beam correction $\text{f}^{p}_{S}$ for
the case where the beam counters do not intercept all the incident beam is:
\begin{equation}
 \text{f}^{p}_{S} = \frac{\lambda_{T}(1-\gamma \text{e}^{-\lambda_{B}})}{
     1-\text{e}^{-\lambda_{B}}}
\label{eqn:poissoncorr}
\end{equation}

To use Eq.~\ref{eqn:poissoncorr}, $\lambda_B$ was obtained in a
straightforward fashion from the observed rate of pions hitting {\em
  both}\/ the BEAM counters {\em and}\/ the target
($\text{R}^{\pi}_{\text{beam}}$). Determination of the rate of pions
traversing the target ($\nu\cdot\lambda_{T}$) had to be inferred from the
observed VETO counter rate ($\nu\cdot(1-\text{e}^{-\lambda_{V}}$)), with
$\gamma$ given as above. The VETO counter was designed to intercept the
entire pion beam in the idealized case of no decay and interaction losses.
These losses were accounted for using our REVMOC \cite{revmoc} beam
simulation to obtain the correct target pion rate and fraction from the
observed VETO rate.  These simulations showed that for $\pi^{+}$ beams,
where the electron contamination was very low, the pion rate at the target
was 97$\pm 0.3$\% of the VETO rate for all energies involved in the
experiment.  The simulation results were confirmed by test runs using a 15
cm diameter scintillator target (the same size as the LH$_{2}$\/ target)
with two different beam intensities, 1 and 2 MHz. For $\pi^{-}$\/ beams,
the pion rate on target varied from about 85\% of the VETO rate at 140 MeV
to about 95\% at 267 MeV.  The multiple pion correction factor,
$\text{f}^p_{S}$, was quite insensitive to the approximations involved in
the REVMOC modelling for the beam intensities involved ($\approx$1 MHz
through the in--beam counters and $\approx$1.5-2 MHz on target).  At these
intensities, a $\pm$5\% variation in the target rate corresponded to
$\delta\text{f}^p_{S}\approx\pm$0.4\%.

The second of the multiple pion correction methods, outright rejection of
such events, also was realized using the VETO counter.  Since the chance of
two pions in a single beam burst {\em both}\/ interacting in the target was
exceedingly small, then if one pion interacted causing a \mbox{$\pi$p}\/
event, the other would pass through the target and be detected by the VETO
counter. Such events were readily eliminated from the \mbox{$\pi$p}\/
yields and the accumulated BEAM hits, B, were corrected appropriately.
Using similar notation to that of the previous section, the appropriate
multiple pion ``veto'' correction factor was f$^{v}_{S}$ = (BEAM events
hitting target with only 1 pion in bucket)/(measured BEAM).  That is:
\begin{equation}
 \text{f}^{v}_{S} = 
 \frac{(\lambda_T\text{e}^{-\lambda_T})\cdot\rho}
 {1-\text{e}^{-\lambda_B}}
\label{eqn:vetocorr}
\end{equation}
Since $\lambda_B$ was known, and $\lambda_T$ and $\rho$ obtained as
described above, Eq.~\ref{eqn:vetocorr} was easily evaluated.

\subsubsection{Test of Multiple Pion Corrections} 
\label{sec:beamratetest}
   
\begin{figure}
\begin{center}
  \mbox{\epsfig{file=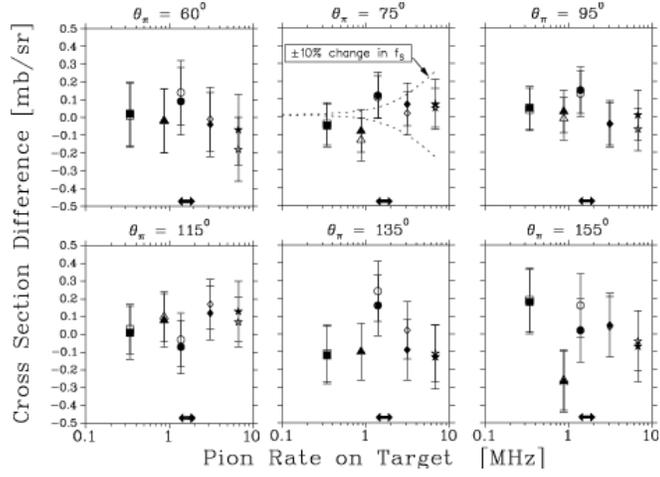, width=8.6cm}}
\end{center}
\caption{$\pi^{+}$p cross section differences for 168.8 MeV  
  coincidence runs using the LH$_{2}$\ target, as a function of incident
  pion rate.  The uncertainties shown are purely statistical.  The
  independent correction methods using the VETO counter (f$^{V}_{S}$ solid)
  and incident BEAM correction (f$^{P}_{S}$ open) differed by at most 0.3\%
  at the highest rate.  The solid double arrow signifies the incident pion
  rate range for production runs. 
  }
\label{fig:ratetest}
\end{figure}

To test our two prescriptions for the multiple pion correction factor,
two-arm \mbox{$\pi^{+}$p}\/ data at 168.8 MeV were obtained for five pion
beam rates on the LH$_{2}$\/ target: 0.34, 0.87, 1.4, 3.1, and 6.7 MHz. One
empty target run with a 1.5 MHz rate was used for background subtraction in
all cases.  Since the beam rate was adjusted using the front end jaws of
the pion channel, the pion beam size on target also changed. However, GEANT
simulations showed that the effect of the jaw changes on the experimental
solid angle was less than 0.2\%.  Consequently, the solid angles
corresponding to the normal beam size were used in all cases. The cross
section results are displayed in Fig.~\ref{fig:ratetest}.  The beam
correction factors varied from 1\% at 0.34 MHz, 4\% at 1.4 MHz, to 26\% at
6.7 MHz.  The maximum difference between the cross sections corrected by
the VETO and Poisson schemes was 0.3\% at 6.7 MHz, and smaller at the lower
rates.  As is evident from this figure, no monotonic systematic variation
within the statistical uncertainties ($\approx$1.3\%) was observed in the
cross sections over this range.  For all two--arm production
\mbox{$\pi^{\pm}$p}\ runs, the cross sections corrected by the Poisson and
VETO schemes never differed by more than 0.3\%, the former being usually
slightly larger than the latter, consistent with expectation based on those
events which should have been vetoed but were not, as discussed in
Sec.~\ref{sec:yields_mult}. The average of the Poisson and VETO corrected
cross sections were used for the two--arm results.
  
However, for the single--arm \mbox{$\pi^{+}$p}\ runs at 141.2, 168.8, and
218.1 MeV, a much larger variation in the cross sections calculated with
the two correction schemes was observed.  The Poisson correction factors
for these runs were the same as those for the two-arm runs (at the same
beam rate and energy), whereas the VETO correction factors were
considerably larger, especially for the target empty runs.  Although the
average discrepancy between the Poisson and VETO corrected cross sections
averaged only about 0.7\%, a systematic dependence with angle was observed,
with the VETO--corrected cross sections 1-2\% smaller than the Poisson
corrected results at 20$^{\circ}$, increasing to be similarly larger than
the Poisson at 70$^{\circ}$.  Since the Poisson scheme relied solely on the
measured beam rates, the resulting correction should have been independent
of pion angle as indeed was observed.  The most likely explanation is that
pion reactions on the background nuclei (mostly carbon) resulted in more
than one charged particle in the final state, with one of them a pion
detected by the TOF Spectrometer, and one of the others a particle detected
by the VETO counter. The fact that the most forward angles were associated
with the largest background is consistent with the most forward angle
yielding the largest discrepancy.  This is also consistent with the
abnormally large correction required for the empty target runs where the
yield is almost solely from reactions on A$>$4 nuclei.  Since the VETO
correction technique was less reliable for these single-arm runs, only the
Poisson correction scheme was used in that case. Such effects were {\em
  not}\/ a concern for the two-arm runs, since in that case the backgrounds
were very much lower, and the coincidence requirement between the pion and
proton arms severely restricted the phase space available for background
reactions yielding another charged particle with a trajectory intercepting
the VETO counter.

\subsection{Yield Extraction} 
\label{sec:yieldextract}

The \mbox{$\pi$p}\/ scattering yield at the pion scattering angle specific
to a particular TOF spectrometer arm, Y(T$_{\pi},\theta_{\text{cm}}$), for
a given number of incident pions onto the target, N$_{\pi}$, was obtained
by accumulating $\pi$2--P1 TOF difference events in the two-arm coincidence
case, or by the $\pi$2 TOF events in the single-arm case.  The desired
\mbox{$\pi$p}\/ scattering yield Y(T$_{\pi}$,$\theta$) in
Eq.~\ref{eqn:dsig} was obtained by subtracting an appropriately normalized
background yield from the foreground, as the foreground yield contained
contributions from both $\pi$p scattering on the hydrogen nuclei as well as
other  pion scattering reactions
on the surrounding material which managed to satisfy the kinematical and
geometrical constraints of the TOF spectrometer. Noting that the incident
pion kinetic energies in the foreground and background runs were within 0.5
MeV in all cases, the $\pi$p scattering yield was determined from:
\begin{eqnarray}
\text{Y} &=& \text{Y}^{\text{fg}} - \biggl(
\frac{\text{N}^{\text{fg}}_{\pi}}{\text{N}^{\text{bg}}_{\pi}} \cdot
\frac{\Delta\Omega^{\text{fg}}}{\Delta\Omega^{\text{bg}}} \cdot
\frac{\epsilon^{\text{fg}}}{\epsilon^{\text{bg}}} \cdot
\frac{\cos\theta_{\text{tgt}}^{\text{bg}}}{\cos\theta_{\text{tgt}}^{\text{fg}}}
\cdot
\frac{\text{N}^{\text{fg}}_{\text{back}}}{\text{N}^{\text{bg}}_{\text{back}}}
\biggl) \cdot \text{Y}^{\text{bg}}_{\text{back}} \nonumber \\ 
 &=& \text{Y}_{\text{fg}} - \kappa\cdot\text{Y}^{\text{bg}}_{\text{back}}
\label{eqn:fgbgnorm}
\end{eqnarray}
where $\text{Y}^{\text{fg,bg}}$ are the yields, $\epsilon^{\text{fg,bg}}$
the efficiencies, $\text{N}^{\text{fg,bg}}_{\pi}$ the number of incident
pions, $\text{N}^{\text{fg,bg}}_{\text{tgt}}$ the areal target densities,
$\Delta\Omega^{\text{fg,bg}}$ the solid angles, and
$\cos\theta_{\text{tgt}}^{\text{fg,bg}}$ the target angles of the
foreground and background targets.

Apart from the obvious dependence on target and beam, the
foreground/background normalization factor,
$\kappa$($\text{T}_{\pi},\theta_{\pi})$, also depended on the pion angle.
This dependence, due to the fact that the effective solid angle
($\Delta\Omega_{\text{eff}}$) differed ($<$1\%) between the foreground and
background targets, arose mostly because of different proton multiple
scattering and/or pion and proton hadronic interaction losses in the two
targets. The differences in the foreground and background solid angles for
the two-arm coincidence setups could be reliably determined by the Monte
Carlo simulation. Also, the particular value of $\kappa$ used in the cross
section calculations also depended on which multiple pion correction scheme
(Poisson or VETO) was used, since they used different beam corrections.

For the two-arm coincidence technique, the foreground TOF difference peaks
were narrow Gaussians with $\sigma\approx 300$ps.  Most of the observed
background under the foreground peak stemmed from pion quasi-elastic
scattering from protons bound in heavier nuclei (mostly carbon) in the
target region. This background was almost negligible for the smaller angles
at 141 MeV, and reached a maximum level of about 7\% for 218 MeV
\mbox{$\pi^{+}$p} reactions in the CH$_2$ target
at backward angles, as illustrated in
Fig.~\ref{fig:fgbgnet}.  The final \mbox{$\pi$p}\/ scattering yield was
determined using wide (several $\sigma$) gates placed around the relevant
spectra.

For the single--arm runs, the candidate \mbox{$\pi^{+}$p}\/ scattering
events were identified by the TOF difference between the $\pi$2 counter and
the BEAM coincidence.  Without the proton arm coincidences to discriminate
against $\pi^{+} + \text{A}\rightarrow\pi^{+} + \text{p} +$X quasi
two--body events, pion quasi-elastic scattering and absorption on carbon
contributed much larger backgrounds than in the two-arm runs. Although the
timing resolution was adequate for separating pions from all but the
fastest protons, the elastically and quasi--elastically scattered pions on
carbon had very similar velocities to those from \mbox{$\pi^{+}$p}\/ and so
could not be separated except via a background subtraction.
Figure~\ref{fig:yield_1arm} is an example of such a $\pi$2 timing spectrum
at the most forward angle (20$^0$).  As for the two-arm case, the
foreground/background normalization factor was determined using
Eq.~\ref{eqn:fgbgnorm}.  However, in the single-arm setup, the pions
scattered in the full LH$_{2}$\/ target could {\em rescatter}\/ in the
surrounding target material (e.g., vacuum vessel, target ring) with the
resulting final state pions or protons detected by the pion arms.  This
served to {\em enhance}\/ the pion arm acceptance when the target was full
relative to when it was empty.  The added acceptance due to pion
rescattering was determined by Monte Carlo simulation as outlined in
Appendix~\ref{appdx:simulDetails}.  The uncertainty in the net yield
arising from the uncertainty in the normalization factor (from the counting
statistics and the uncertainty in the relative target angle between
foreground and background runs) was added in quadrature to the other
statistical uncertainties. Despite the sizable backgrounds, the
foreground--background subtraction resulted in a clean \mbox{$\pi^{+}$p}\/
yield peak. An estimated 0.2$\pm$0.2\% residual proton background (from the
adjacent peak) was subtracted from the yields. Since the positions of the
background peaks were often a few channels different from those of the
foreground (due to small energy loss differences, timing drifts), the peaks
were shifted appropriately before subtraction. In practice, such shifts
made no statistical difference to the yields. The extracted net yields were
defined by a software gate placed around the \mbox{$\pi^{+}$p}\/ peak.
Slightly wider/narrower gates were used to check the sensitivity to the cut
placement, and any differences were included in the overall statistical
uncertainty.
 
\subsubsection{Yields and Multiple Pion Correction}
\label{sec:yields_mult}

When dealing with the multiple pion events, due care had to be directed to
the yield definition as well as to the beam normalization. In the Poisson
correction scheme, the number of incident pions detected by the in--beam
counters was {\em increased}\/ by the factor in Eq.~\ref{eqn:poissoncorr}
to account for the multiple pion events.  In this scheme, the $\pi$2-P1 TOF
difference spectra included contributions from both single and multiple
pion BEAM events, so no additional constraints needed to be applied to the
yield spectra.

In the VETO correction scheme, a multiple pion event was identified by a
particle hitting the VETO paddle at the same time as detection of a
$\Pi\cdot$P coincidence in one of the TOF spectrometer arms.  These events
were then removed from the $\pi$2-P1 TOF difference spectra, resulting in a
yield corresponding to only single--pion BEAM events.  Correcting the
resulting incident beam N$_{\pi}$ for these rejected events using
Eq.~\ref{eqn:vetocorr} resulted in cross--section values appropriate to
single--pion beam events.  In this (VETO) correction scheme, however, two
special cases had to be considered, one where an event was vetoed but
shouldn't have been, and vice versa.

In the first case, the extra particle(s) in the beam bursts could have been
muons or electrons, which passed through the target and hit the VETO
counter, while a pion in the same burst caused the $\Pi\cdot$P event.
Although these \mbox{$\pi$p}\/ events were rejected using the VETO cut, the
incident BEAM count was also corrected for such events, so the effect
cancels, the net result being simply a loss of statistics.

In the second case, where two (or more) pions were incident on the target
in a beam burst, with one causing a $\Pi\cdot$P event, and the other
continuing on to the VETO paddle, the latter pion could have interacted
(e.g.\ decayed) prior to reaching the VETO counter with the interaction
products (e.g., decay muon) {\em escaping}\/ detection by the VETO\@. Here,
the events should have been rejected but were not.  Results from GEANT
simulations showed that of the beam pions traversing the target, only about
6\% failed to cause a VETO hit by either the pion or its decay muon.  Thus
for a typical 3\% multiple pion correction, only about 0.2\% of the events
should have been identified as multiple pion events, but were not, a small
effect consistent with observation.

 
\section{Results} 
\label{sec:results}

The many systematic checks that were performed to test our determinations
of the effective solid angles, target thicknesses, and beam normalization
(as illustrated in Figs.~\ref{fig:angletest}, \ref{fig:syst_checks},
\ref{fig:thicktest}, \ref{fig:prt_fractions}, \ref{fig:ratetest}),
indicated that the system was well understood. In all cases the test results
were well within the normalization uncertainties ascribed to them. In fact,
for the case of the LH$_2$ target angle, the test data (along with the
overlapping $\pi^+$p  single--arm and two--arm data) was {\em essential}\/
to help determine the magnitude of the systematic angle offset and to
estimate its uncertainty.

\begin{figure}
\begin{center}
  \epsfig{file=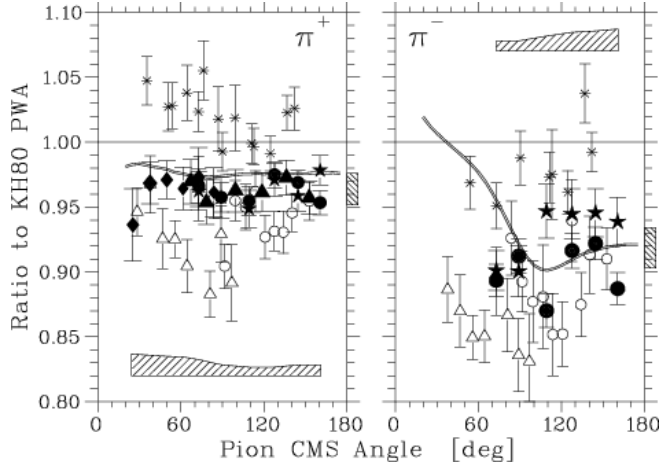, width=8.6cm}
\end{center}
\caption{
  Differential cross section results at 141.15 MeV ($\pi^+$p left and
  $\pi^-$p right) plotted as ratios to the KH80 PWA
  solution\protect\cite{kh80}. The solid diamond, circle, triangle, and
  star points represent the single--arm LH$_2$, two--arm LH$_2$ set ``A''
  and set ``B'', and two--arm CH$_2$ results, respectively.  The error bars
  shown are statistical only.  The results of Bussey et
  al.\protect\cite{bus73} (asterixes) and Brack et
  al.\protect\cite{bra86,bra95} (open points) are shown as ratios to KH80
  at their respective energies.  The double line is the prediction of the
  V.P.I. SM95 PWA\protect\cite{sm95}\@.  The horizontal hatched area
  represents the undertainty due to a 1$\sigma$ {\em increase}\/ in the
  beam energy, and the vertical hatched areas represent the typical $\pm
  1\sigma$ normalization uncertainties for {\em each}\/ of our data sets.
  \normalsize }
\label{fig:dsgratios141}
\end{figure}

\begin{figure}
  \begin{center}
    \mbox{\epsfig{file=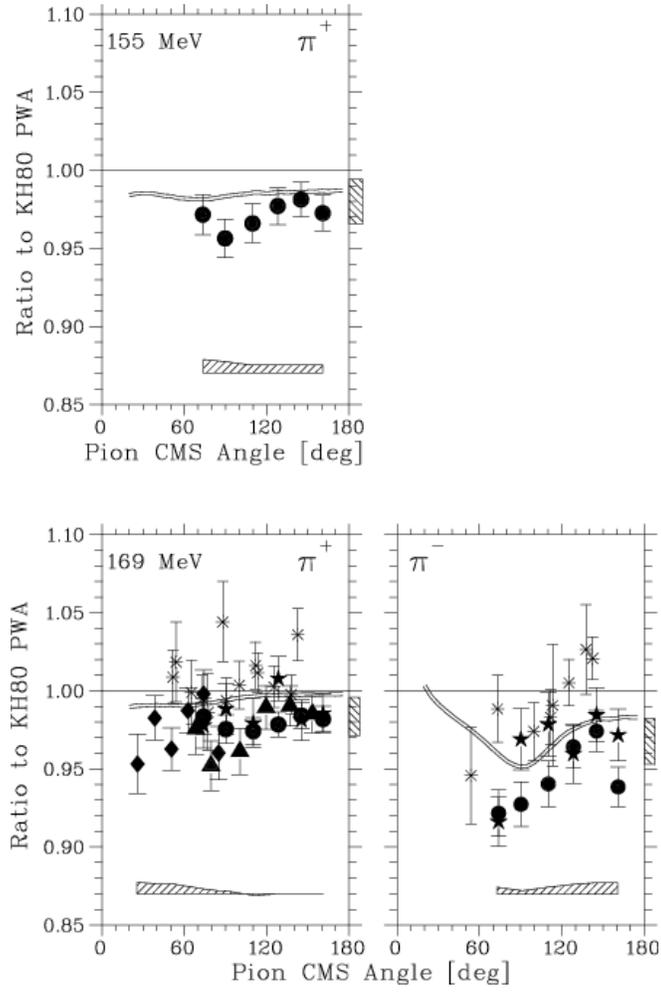, width=8.6cm}}
  \end{center}
  \caption{
    As Fig~\protect\ref{fig:dsgratios141}, but showing our 154.6 MeV
    $\pi^+$p data (top) and our 168.8 MeV $\pi^{+}$p and $\pi^{-}$p data
    (bottom). Note that at 168.8 MeV, the $\pi^+$ +1$\sigma$ error band due
    to the energy uncertainty changes sign near 100$^{\circ}$ then stays
    near zero.}
  \label{fig:dsgratios155-169}
\end{figure}

\begin{figure}
\begin{center}
\mbox{\epsfig{file=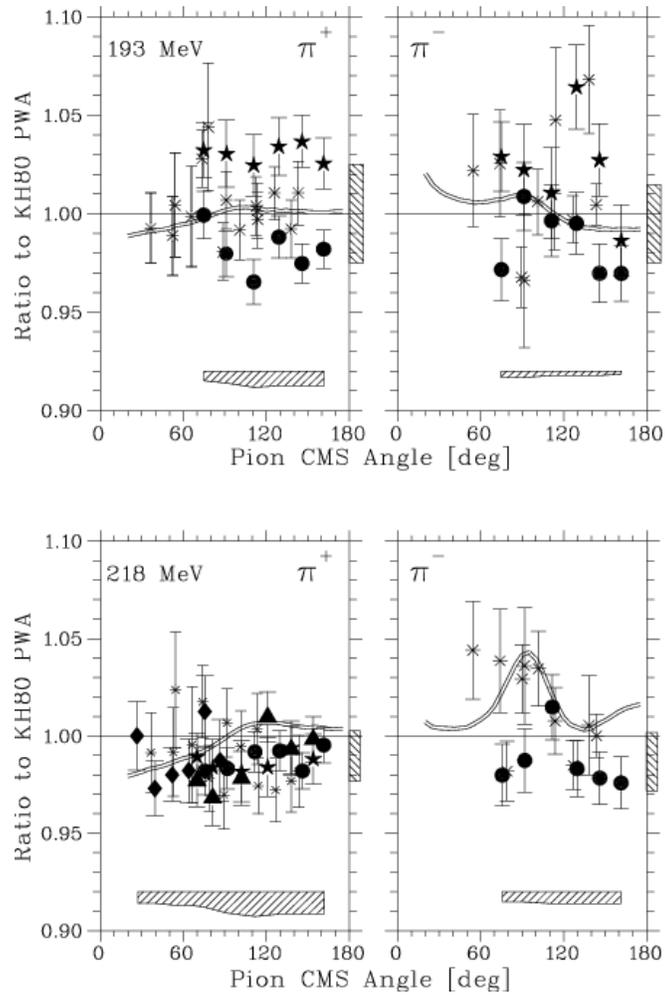, width=8.6cm}}
\end{center}
\caption{ 
  As Fig.~\protect\ref{fig:dsgratios141}, but showing our 193.2 MeV
  $\pi^{+}$p and $\pi^{-}$p data (top) and our 218.1 MeV $\pi^{+}$p and
  $\pi^{-}$p data (bottom).  Note that at 218.1 MeV and higher energies,
  the +1$\sigma$ error band due to the energy uncertainty has changed sign
  with respect to Fig.~\protect\ref{fig:dsgratios141}.  }
\label{fig:dsgratios193-218}
\end{figure}

Both single--arm and two--arm $\pi^+$p scattering data were obtained at
141.2, 168.8, and 218.1 MeV, as shown in Figs.~\ref{fig:dsgratios141},
\ref{fig:dsgratios155-169}, and~\ref{fig:dsgratios193-218}.  In general the
agreement between the single-- and two--arm results in their angular region
of overlap is excellent.  The Set ``B'' $\pi^+$p scattering data were
obtained at 141.2, 168.8, 218.1, and 267.3 MeV (as shown in
Figs.~\ref{fig:dsgratios141}, \ref{fig:dsgratios155-169},
\ref{fig:dsgratios193-218} and~\ref{fig:dsgratios241-267}) in the middle of
the experimental running period, and necessitated removing and
repositioning the TOF Spectrometer pion arms.  The target angle was
slightly different as well, 50.6$^{\circ}$ compared to 53.6$^{\circ}$. The
excellent agreement of the Set B data with the corresponding Set A data
provides additional confirmation of the positional accuracy of the counter
arms and the accuracy of our solid angle determinations.

Data were obtained using both solid and liquid targets for
\mbox{$\pi^{+}$p}\/ scattering in the two--arm configuration at 141.2,
168.8, 193.2 and 218.1 MeV and for \mbox{$\pi^{-}$p}\/ scattering at 141.2,
168.8 and 193.2 MeV (as shown in Figs.~\ref{fig:dsgratios141},
\ref{fig:dsgratios155-169}, and~\ref{fig:dsgratios193-218}). In general,
the agreement between the solid and liquid target results is good, within
the ascribed normalization uncertainties, {\em except}\/ at 193.2 MeV,
where the solid target results were consistently larger than the liquid
target results for both \mbox{$\pi^{+}$p} and \mbox{$\pi^{-}$p}.  Neither
the beam normalization constants (f$^{\pi}$, f$_S$, etc.), the foreground
and background yields, the effective solid angles, nor the beam energy in
these sets appear out of line with respect to the adjacent energies, so the
source of the discrepancy is unclear.  Possible reasons include erroneous
settings of the CH$_2$ target angles (which were readjusted for each
energy) and/or the momentum selecting slits, which if mistakenly adjusted
off centre, could cause the central beam energy to shift.  Neither of these
would have caused any change in our diagnostics or our data and thus would
have escaped detection. Since at every other energy the agreement among the
various experimental configurations is good, the normalization
uncertainties for just these four data sets at 193.2 MeV were increased, to
2\% for the two \mbox{$\pi^{-}$p}\/ runs, and 2.5\% for the two
\mbox{$\pi^{+}$p}\/ runs, to bring them into agreement at the limit of
their 1$\sigma$ normalization uncertainties.

\subsection{The Absolute Differential Cross Sections} 
\label{sec:absdiffxsect}

The final results for the \mbox{$\pi^{+}$p}\/ and \mbox{$\pi^{-}$p}\/
elastic absolute differential cross sections in the centre--of--mass system
are listed in Tables~\ref{tab:dsg_results141}
through~\ref{tab:dsg_results267}. The uncertainties quoted are the usual
1$\sigma$ values.  Common uncertainties such as those associated with beam
energy and normalization are {\em not}\/ included in the errors associated
with each data point, but are listed separately.  All the data of each type
were obtained from runs characterized by a fixed experimental configuration
(i.e. beam rate, target angle, etc.), except for the 168.8 MeV
\mbox{$\pi^{+}$p}\/ LH$_{2}$\/ and CH$_{2}$\/ two-arm results, which are
weighted averages of runs taken with three different beam rates, and three
different target thicknesses, respectively.  The justification for this
averaging is provided by Figs.~\ref{fig:thicktest} and~\ref{fig:ratetest},
which indicated {\em no}\/ systematic dependence of the cross-section on
these parameters.  The final cross sections had statistical uncertainties
of $\sim$1-1.5\% for $\pi^+$p and $\sim$1.5-2\% for $\pi^-$p, each with
$\sim$1-1.5\% normalization uncertainties.

\subsubsection{Uncertainties in the Absolute Normalization}
\label{sec:errors}
   
The normalization uncertainties quoted in the tables are based on the
following considerations:

\begin{itemize}
  
\item {\bf Target Angle, cos$\theta_{\text{tgt}}$:}
  $\pm$0.4$^{\circ}$(zero offset)$\pm0.2^{\circ}$(reproducibility) for the
  LH$_{2}$\/ target, corresponding to a $\pm$1.1 (1.0)\% uncertainty in
  cos$\theta_{\text{tgt}}$ for $\theta_{\text{tgt}}$ = 53.6
  (50.6)$^{\circ}$ in the two-arm setups, and $\pm$0.6\% for
  $\theta_{\text{tgt}}$ = -39.4$^0$ (single-arm setup). For the CH$_{2}$\/
  targets, we estimate an uncertainty of $\pm$0.25$^0$ corresponding to
  $\delta$cos$\theta_{\text{tgt}}$= 0.6\%.  These estimates are based on
  the results discussed in Sec.~\ref{sec:angletest}.
\item {\bf Multiple Pion Correction, f$_{S}$:} A conservative estimate of
  $\pm$10\% is ascribed to the value of $|1-\mbox{f}_{S}|$ determined for
  each run, a value justified by the excellent agreement exhibited by the
  results discussed in section~\ref{sec:beamratetest} and shown in
  Fig.~\ref{fig:ratetest}.
\item {\bf Pion Fraction, f$_{\pi}$:} For the \mbox{$\pi^{+}$p}\/ data the
  uncertainties ranged from $\pm$0.3\% at 141.1 MeV to $\pm$0.1\% at 267.2
  MeV, as inferred from direct measurements during the phase-restricted
  beam operation described in Sec.~\ref{sec:beamcomp} and shown in
  Fig.~\ref{fig:prt_fractions}.  For the \mbox{$\pi^{-}$p}\/ data, the
  uncertainties ranged from 0.9\% at 141.1 MeV to 0.3\% at 267.2 MeV.  Up to
  193 MeV, the uncertainties are associated with the fits to the TCAP
  spectra as discussed in  Sec.~\ref{sec:beamcomp}.
\item {\bf Pion Decay, f$_{D}$:} $\pm$0.2\% in all cases, since the results
  of the GEANT and REVMOC simulations used to generate these corrections
  agreed to $<$0.1\%.  Another 0.1\%
  was added for the uncertainty in the contribution from pion decay within
  the channel as discussed in Sec.~\ref{sec:beamdecay}. 
\item {\bf Hadronic Interaction Loss, f$_{L}$:} This uncertainty was
  estimated to be 15\% of the calculated loss to the centre of the target,
  varying from 0.3\% for 168.8 MeV \mbox{$\pi^{+}$p}\/ on
  the LH$_{2}$\/ targets, to 0.1\% for 193.2 MeV \mbox{$\pi^{-}$p}\/ on the
  2 mm CH$_{2}$\/ target. 
\item {\bf Target Proton Density, N$_{\text{prot}}$:} The uncertainty in the
  proton density of the LH$_{2}$\/ target, $\pm$0.5\%, was estimated from
  the vapour bulb measurements conducted during the experiment (see
  section~\ref{sec:lh2target}).  The uncertainty in the proton density
  of the CH$_{2}$\/ targets was $\pm$1\% , as measured by chemical analysis
  by a commercial laboratory \cite{brack-thesis} (see
  Sec.~\ref{sec:ch2target}). 
\item {\bf Beam and Computer Live Time, B and f$_{\text{LT}}$:} The
  uncertainties in both these quantities were negligible,
  $<$0.1\%, since the particle counting was done with several independent
  scaler modules with no discrepancies observed. The live-time f$_{\text{LT}}$
  was 0.98 or better for the data presented in the tables.
\end{itemize}

All of the normalization uncertainties outlined above were combined in
quadrature to yield the values quoted in the tables.  
As an example, Table~\ref{tab:typerrs} shows the uncertainties and 
their sum for the 168.8 MeV data.

\subsubsection{Angle--Dependent Uncertainties}
   
The experimental uncertainties which depended on the pion scattering angle
are the counting statistics in the foreground and background
runs, the statistics in the Monte Carlo determinations of the solid angles,
the uncertainties in the hadronic loss corrections of the scattered pions
and recoil protons, and the uncertainty in the distance from the relevant
$\pi$2 counter to the target centre ($\pm$0.5\% corresponding to $\pm$3
mm).
   
The uncertainty in the net yield Y given by Eq.~\ref{eqn:fgbgnorm} is $
\Delta\text{Y} = \sqrt{(\delta\text{Y}^{\text{fg}})^{2} +
  (\kappa\cdot\text{Y}^{\text{bg}}_{\text{back}})^2 \cdot
  \bigl((\frac{\delta\kappa}{\kappa})^2 +
  (\frac{\delta\text{Y}^{\text{bg}}_{\text{back}}}{
    \text{Y}^{\text{bg}}_{\text{back}}})^2 \bigl) }$ where the
uncertainties in the foreground and background yields are Poisson
distributed and the foreground/background normalization uncertainty
$\delta\kappa$ arises mainly from the target angle uncertainties in the
foreground and background runs.  In practice, the second term involving
$\kappa$ was negligible in the two-arm measurements where the backgrounds
were very small, but it was non--negligible in the single-arm runs, where
backgrounds were typically 25\% (but even up to 50\% at 20$^0$) of the
foreground yields.  Also, for the single-arm runs, there was some
uncertainty in the yields arising from the placement of the software cuts,
uncertainties which were added in quadrature to the other uncertainties. In
practice, these variations were never larger than half of the statistical
uncertainties. The hadronic loss uncertainty was estimated to be 10\% of
the actual loss suffered by the pions and protons.  The final
``statistical'' uncertainties were obtained by summing in quadrature all
these separate components. Reference~\cite{mythesis} provides a sample
calculation of all solid angle corrections.

\subsubsection{Radiative Corrections}
 
In experiments utilizing magnetic spectrometers to detect scattered charged
particles, the fraction of events lying outside the spectrometer energy
acceptance due to bremsstrahlung energy loss would have to be considered.
In our measurements, however, the times--of--flight of the pions and
protons were measured, not the energy, and so energy losses would manifest
themselves as tails in the timing distributions.  No such tails were
observed in any of our spectra.  As the cross section for the
bremsstrahlung process ($\pi p\rightarrow \pi p \gamma$) is known to be
very small ($<<$0.1 mb/sr) \cite{fearing}, any radiative corrections would
have been negligible compared to the other uncertainties characterizing the
experiment. Consequently, no radiative corrections were applied to the
data.

 
\section{Discussion} 
\label{sec:conclusion}

\begin{figure}
\begin{center}
\mbox{\epsfig{file=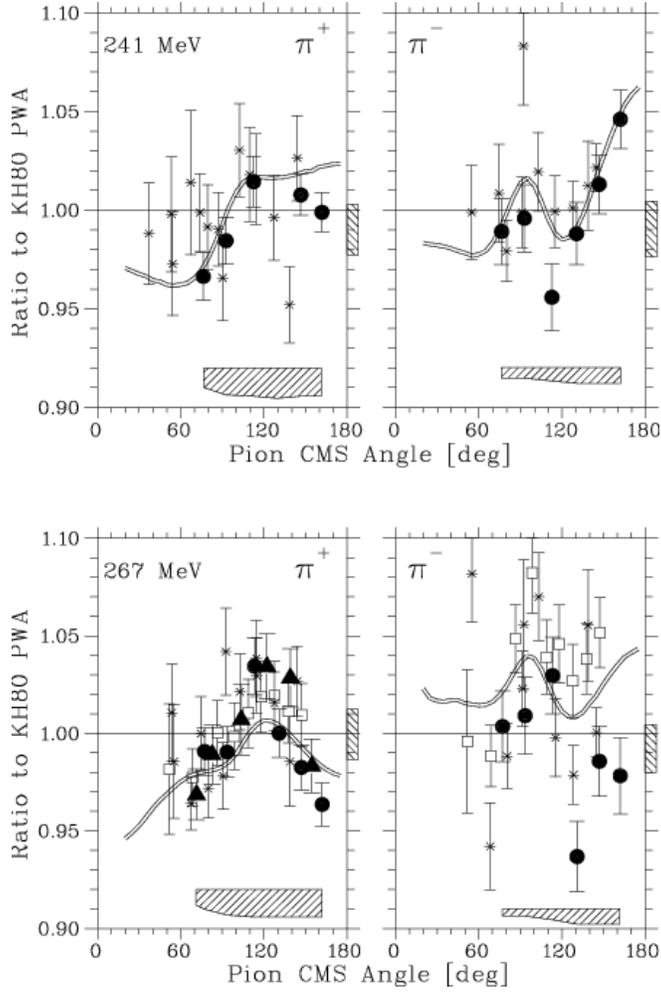, width=8.6cm}}
\end{center}
\caption{
  As Fig.~\protect\ref{fig:dsgratios141}, but showing our 240.9 MeV
  $\pi^{+}$p and $\pi^{-}$p data (top) and our 267.3 MeV $\pi^{+}$p and
  $\pi^{-}$p data (bottom), where the LAMPF data of Sadler, et al.\ (263
  MeV)\protect\cite{sad87} are also shown (open boxes).}
\label{fig:dsgratios241-267}
\end{figure}

In Figs.~\ref{fig:dsgratios141} through~\ref{fig:dsgratios241-267}, the
cross section results are shown as ratios to the Karlsruhe--Helsinki KH80
PWA solution \cite{kh80}.  Also shown are the results of the last published
PWA from the V.P.I. group, SM95 \cite{sm95} and the data of Bussey et al.
\cite{bus73}, Brack et al. \cite{bra86,bra95}, and Sadler et al.
\cite{sad87}, all plotted as ratios to KH80 at their respective energies.
The use of such ratios enables meaningful comparisons since the data sets
were measured at somewhat different  energies, and also highlights
differences between the data sets which would not be visible on an absolute
scale.

Prior to this work, the results of Bussey et al. \cite{bus73} constituted
the only comprehensive set of differential cross sections for energies
spanning the $\Delta$ resonance. The two previous TRIUMF experiments of
Brack et al. \cite{bra86,bra95} covered a range of energies up to $\sim$139
MeV, whereas those of the LAMPF group of Sadler et al. \cite{sad87} were at
higher energies, extending down to 263 MeV, both of which overlap our
energy range. Although there are also $\pi^+$p data up to 140 MeV by
Ritchie et al \cite{ri83}, the 140 MeV data were not included in
Fig.~\ref{fig:dsgratios141} for reasons of clarity\footnote{Other data sets
  \protect\cite{otherData} also were excluded in figures
  \protect\ref{fig:dsgratios141} through \protect\ref{fig:dsgratios241-267}
  due to the few energies covered and the large error bars associated with
  the data, which limits their impact in partial wave analyses.}. Of
particular interest are the results of Brack et al.\ which employed a
spectrometer similar to that used in our experiment.  Whereas their first
experiment \cite{bra86} used solid CH$_{2}$\/ targets in a $\pi p$ two-arm
coincidence configuration, their second \cite{bra95} used an active
scintillator target to detect the recoil proton. The pion arm scintillators
used in their experiment were also different from ours.

Our lowest energy results agree within uncertainties with the earlier
two-arm coincidence measurements of Brack et al.\ \cite{bra86}, although
the latter are systematically lower than ours by 1--2\%. The results of the
forward angle active target experiment of Brack et al.\ \cite{bra95} are
also consistent with our data, although at the edge of the relative
normalization uncertainty of about 3\%. Our $\pi^+$p results at 141 MeV are
also in good agreement with the data of Ritchie et al. \cite{ri83} at 140
MeV.  Our highest energy results at $\approx$267 MeV are completely
consistent in both normalization and shape with those of Sadler, et al.
\cite{sad87} which have 3\% and 5\% normalization uncertainties for the
\mbox{$\pi^{+}p$}\/ and \mbox{$\pi^{-}p$}\/ data respectively. Comparison
of our data with those of Bussey et al.\ yield a mixed picture\footnote{We
  refer to the $\approx$1\% normalization uncertainties ascribed to the
  Bussey data after publication \protect\cite{bugg-renorm}\/, and not the
  larger 5\% value adopted in the SM95 analysis \protect\cite{sm95}. No
  value was quoted in the original paper\protect\cite{bus73}.}. Above the
resonance, there is consistency within the stated uncertainties. However at
energies below the resonance peak, our results are systematically lower
than theirs, particularly for $\pi^{-}p$, with the largest disagreement
occurring at 141 MeV. It is noteworthy that at the lower energies, the
Bussey et al. \cite{bus73} cross-sections are systematically larger than
the Brack et al. \cite{bra86,bra95} results, with our data slightly below
half-way between the two sets at 141 MeV. Also, our data have better
statistical precision than any of these other data sets.

Our results are systematically lower than, and in clear disagreement with,
the KH80 PWA solution at energies below the resonance, though there is
better agreement with KH80 above the resonance. Our results agree rather
well with the predictions of the SM95 solution at all energies, even though
our data were not used in the SM95 fitting. Note that the Bussey et
al.\ \cite{bus73} data were included in the KH80 database with {\em no}\/
normalization uncertainties, but were assigned 5\% normalization
uncertainties in the SM95 database (compared to the 0\% quoted \cite{bus73}
and the 1\% provided subsequently \cite{bugg-renorm}) in order to resolve
an inconsistency between the single-energy and global fits in the SM95
solution.  The increased uncertainty, hence reduced weight, of the Bussey
et al., data resulted in a solution more consistent with the global
database. The generally good agreement between our results and the SM95
predictions demonstrate that our results are more compatible with the
global database in the vicinity of the Delta resonance than were the
previous measurements. In addition, we have provided data with superior
statistical precision. The precision is such that shape differences due to
the difference in the small D-waves between the SM95 and KH80
solutions are clearly seen in the data at the highest energies.

Although the full impact of our data can only be appreciated after a new
global PWA fit\footnote{Preliminary solutions which include our data can be
  found at the SAID site \protect\cite{said}}, the consistency between our
data and the SM95 PWA solution permit some preliminary observations. The
SM95 solution provides a good fit to the total cross section results of
Pedroni et al.\ \cite{ped78}, but much less so to those of Carter et al.\ 
\cite{car71} (which are larger than Pedroni on the left-wing of the
resonance), so our data support the former data set.  Our data also support
the value of the $\pi$NN coupling constant derived from the SM95 solution,
$f^2 \sim$0.076, over that of KH80 (0.079), a result which can be traced to
the smaller mass and narrower width of the Delta resonance in the SM95
solution compared to KH80. The lower value of the coupling constant is
known to resolve long standing inconsistencies in the Goldberger-Treiman
discrepancy \cite{gol58} and Dashen-Weinstein sum rule
\cite{das69,fuc94,goit99}. The $\pi$N sigma term from SM95 and subsequent
preliminary solutions which include our data \cite{pav99} seem to indicate
a sigma term {\em larger}\/ than the canonical result \cite{koch82}, thereby
{\em increasing}\/ the discrepancy with the theoretical result \cite{gas81}.
However, since the sigma term is a very difficult quantity to extract from
the $\pi$N data, definitive conclusions cannot be reached at this time.

\subsection{Concluding Remarks}
\label{sec:discussion}

The primary goal of our experiment was to provide absolute differential
cross section data with small and {\em reliable}\/ uncertainties at
energies spanning the $\Delta$ resonance.  In addition, the results of many
test measurements proved invaluable in elucidating the nature of systematic
errors which were subsequently corrected or accounted for in the
uncertainties.  The satisfying internal consistency of our results
demonstrates that our goal of obtaining reliable estimates of systematic
uncertainties was successfully attained.  It also demonstrates that the
techniques employed in this experiment and shared by the previous work of Brack
et al. \cite{bra86,bra95,bra88}, such as the use of solid targets, were
reliable, despite criticisms to the contrary (see e.g. \cite{buggCrit}).

These data resolve a long standing controversy regarding the inconsistency
of cross section values around the Delta resonance with those at lower
energies. Consequently, inclusion of the results of this work will result
in a more consistent database than was previously available.  As a result
of the improved consistency in the database and the increased precision of
these data over previous results, more accurate and reliable determinations
of the $\pi$NN coupling constant and $\pi$N sigma term can be expected.

\section{Acknowledgements}

The authors wish to acknowledge the National Science and Engineering
Council of Canada for financial support, the TRIUMF cryogenic target and
cyclotron operation groups for their invaluable assistance, and PNPI
Gatchina for providing their surface barrier detectors. M.M.P. acknowledges
support by the Wescott Fellowship and E.W. Vogt, and thanks G. Stinson and
R.A. Pavan for assistance with the REVMOC and TRANSPORT beam transport
codes.  I.S. acknowledges the hospitality of TRIUMF, the support of E.W.
Vogt, and the NATO Collaborative Research Grant \#921155U.


 
\appendix

\section{Hadronic Interaction Corrections to the Solid Angles}
\label{appdx:simulDetails}
\label{sec:hadrcorr}

The GEANT~\cite{geant} simulations described in Sec.~\ref{sec:geantsldang}
included all the geometrical constraints and physical processes that could
affect the solid angle, except for pion or proton inelastic interactions
with nuclei, for which GEANT was found to be unreliable in the energy
region involved in this experiment\footnote{As GEANT v3.21 was designed as a
  simulation tool for high energy physics, many of the approximations used
  in the hadronic interaction routines are not appropriate at low
  energies.}. Hadronic elastic and quasi-elastic interactions
were included to simulate pion hadronic rescattering into the pion arms
from the cryostat vessel and target ring.  However, as the hadronic
rescattering corrections were never large, but only up to 1\% in the
forward angle single arm runs, even a large uncertainty there would not
increase the overall solid angle uncertainty appreciably. In practice, an
additional uncertainty of 33\% of the additional correction was applied in
those cases.

\begin{figure}
\begin{center}
  \mbox{\epsfig{file=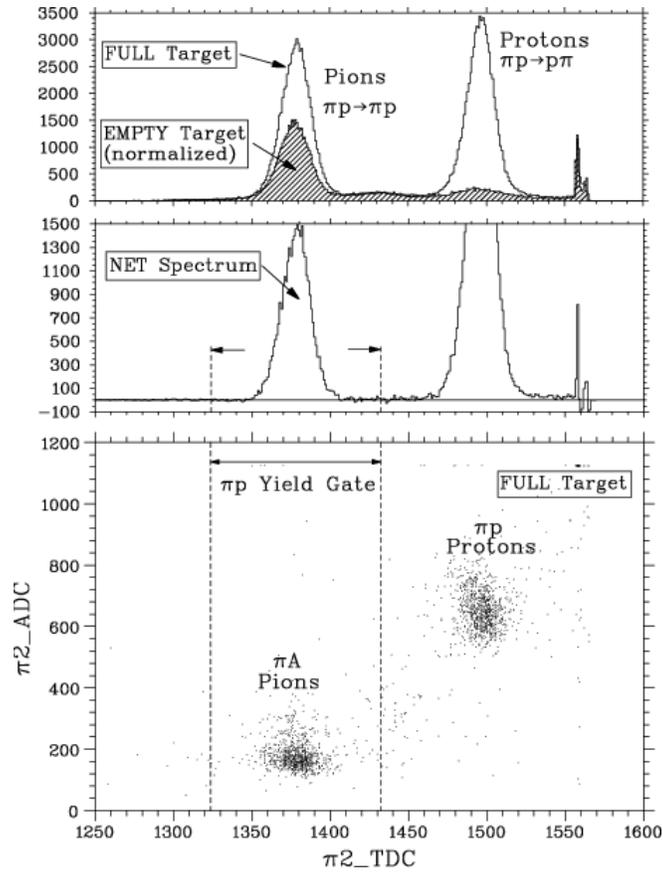, width=8.6cm}}
\end{center}
\caption{
  {\bf Top:} A $\pi$2 timing spectrum ($\theta_{\pi}$=20$^{\circ}$) showing
  the software gate used to extract \mbox{$\pi p$}\ yield from a 169 MeV
  $\pi^{+}$p single--arm run. The background level was maximal here and
  decreased with larger angle.  The protons shown, corresponding to
  backward going pions, were fast enough to satisfy the hardware EVENT
  timing requirement. {\bf Bottom:} $\pi$2 foreground pulse height versus
  timing spectrum, showing clear separation of pions and protons. The
  $\pi$p yield was defined using the one dimensional gate shown.}
\label{fig:yield_1arm}
\end{figure}

\begin{figure}
\begin{center}
  \mbox{\epsfig{file=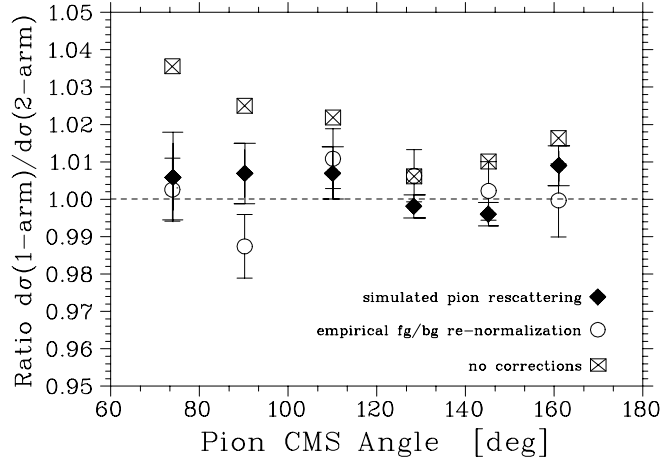, width=8.6cm}}
\end{center}
\caption{
  Ratio of the $\pi^{+}$p single--arm cross section to the coincidence
  values at 169 MeV using the LH$_{2}$ target, for a run where the two were
  obtained {\em simultaneously}.  The filled diamonds (crossed boxes) show
  the results where the corrections due to pion hadronic rescattering in
  the vacuum vessel were (were not) included in the solid angle
  simulations.  The circles show the result using the uncorrected solid
  angles, where the full--/empty-- target normalization was obtained by
  matching the backgrounds.  The uncertainties reflect the 33\% simulation
  uncertainty in the additional rescattering correction, or the
  uncertainties in the full/empty normalization. The corrected results
  verify that the rescattering correction was under control.}
\label{fig:lastfigure}
\end{figure}

A specific test was carried out to check these rescattering corrections. In
this test, the proton arms were removed from the EVENT coincidence. Though
now in single--arm detection mode, the proton arm information was still
recorded, so both single-arm and two-arm yields were obtained {\em
  simultaneously}\/. Comparison of the single-arm and two-arm
cross-sections {\em without}\/ including pion hadronic rescattering in the
vacuum vessel and target ring showed that the single--arm results were
systematically larger than the two--arm results by about 1 to 4\%
(Fig.~\ref{fig:lastfigure}). However, including the additional pion hadronic
rescattering contribution brought the two into agreement.  As an additional
check, another empirical method was used to account for the additional
rescattering acceptance (see Ref.~\cite{bra88}). The ratio between a full
and empty target run of counts in a software box placed around the
background pulse--height versus timing spectrum (e.g.
Fig.~\ref{fig:yield_1arm}) was used for the foreground/background
normalization constant (see Eq.~\ref{eqn:fgbgnorm}). Additional
contributions to empty target background from pion {\em quasi}-elastic
rescattering in the vacuum vessel appear in the background software box,
and this increase closely mimics the increase of the {\em elastic}\/
rescattering in the true yield box.  Uncertainties arise from the counting
statistics in the software boxes, and from the fact that the quasi-elastic
rescattering contribution does not necessarily match exactly the elastic
contribution. The additional uncertainties are not expected to be bigger
than about a third of the total rescattering correction.  The results from
this method also are shown in Fig.~\ref{fig:lastfigure}, and are seen to
agree with the simulated hadronic rescattering corrections.

Unlike the pion (quasi-)elastic rescattering corrections, the hadronic
inelastic interaction {\em losses}\/ of the pions and protons through the
target, air (and for pions, the $\pi$1 counter) on their way to their
respective counters were relatively large (up to 7\%). A program was
developed to calculate these losses, since it was found that the GEANT
simulations were unreliable. Consequently, the hadronic losses were
determined using the program and then applied {\em post priori}\/ to the
GEANT solid angle results calculated with hadronic inelastic interaction
disabled. Details, including a sample loss calculation, can be found in
Ref.~\cite{mythesis}.


\newpage

\newpage

\newpage

\newpage

\begin{table}
\caption{
 Areal densities of target nuclei for the carbon, LH$_{2}$\ and CH$_{2}$\
 targets used in the experiment.
 \label{tab:target_thick}
  }
\begin{tabular}{cr@{$\pm$}lr@{$\pm$}l}
Target & \multicolumn{2}{c}{Thickness} & \multicolumn{2}{c}{H Thickness} \\
       & \multicolumn{2}{c}{[mg/cm$^2$]} & 
                               \multicolumn{2}{c}{[10$^{-6}$ mb$^{-1}$]} \\
\tableline
LH$_{2}$\       & 106.2 &  0.5 & 63.43 &  0.32 \\
CH$_{2}$\ ``A'' & 44.0  &  0.1 &  3.78 &  0.04 \\
CH$_{2}$\ ``D'' & 185.8 &  0.7 & 16.00 &  0.16 \\
CH$_{2}$\ ``E'' & 294.2 &  0.3 & 25.30 &  0.25 \\
graphite      & \multicolumn{2}{c}{285.7} & - & - \\
\end{tabular}
\end{table}

\begin{table}
\caption{ 
 Normalization (i.e.\ scattering angle independent) and typical
 angle--dependent uncertainties (100$\cdot\sigma_X/X$) for the 168.8 MeV
 cross sections. All columns except the first refer to the two-arm
 configuration. The uncertainties at other energies are similar. The
 quantities in parentheses indicate final results which  were obtained  by
 averaging several runs (see text). The normalization  factors are  defined
 in equations \protect\ref{eqn:dsig} and  \protect\ref{eqn:dsigbeam}.
 }
\label{tab:typerrs}
\begin{center}
\begin{tabular}{lccccc}
 \multicolumn{6}{c}{NORMALIZATION  UNCERTAINTIES (\%)} \\
\tableline
 & \multicolumn{3}{c}{\mbox{$\pi^{+}p$}\ } &
            \multicolumn{2}{c}{\mbox{$\pi^{-}p$}\ } \\ \cline{2-4} \cline{5-6}
Factor   & 1 Arm LH$_2$ & LH$_{2}$\ & CH$_{2}$\ ``D'' & LH$_{2}$\ & CH$_{2}$\ ``D'' \\
\tableline
N$_{\text{prot}}$ & 0.5    & 0.5     & 1.0    & 0.5     & 1.0 \\
cos $\theta_{\text{tgt}}$ & 0.6 & 1.1 (0.9) & 0.6 & 1.1 & 0.6 \\
B                 & $<$0.1 & $<$0.1  & $<$0.1 & $<$0.1  & $<$0.1 \\
f$_{\pi}$         & 0.2    & 0.2     & 0.2    & 0.6     & 0.6 \\
f$_{D}$    & 0.2    & 0.2     & 0.2    & 0.2     & 0.2 \\
f$_{L}$    & 0.3    & 0.3     & 0.2    & 0.2     & 0.1 \\
f$_{S}$    & 0.4    & 0.4 (0.1) & 0.4  & 0.6     & 0.6 \\
f$_{\text{LT}}$   & $<$0.1 & $<$0.1  & $<$0.1 & $<$0.1  & $<$0.1 \\
edge effect       & 0.1    & 0.1     & 0.1    & 0.1     & 0.1 \\
\tableline
 $\sqrt{\sum(\Delta_{i})^2}$ & 1.1 & 1.3 (1.1) & 1.3 & 1.4 & 1.5 \\
\tableline
 & & & & & \\
 \multicolumn{6}{c}{ANGLE DEPENDENT UNCERTAINTIES (\%) } \\
\tableline
 & \multicolumn{3}{c}{\mbox{$\pi^{+}p$}\  } & 
            \multicolumn{2}{c}{$\pi^-$p} \\ \cline{2-4} \cline{5-6}
Source & 1 Arm LH$_2$ &  LH$_{2}$\ & CH$_{2}$\ ``D'' & LH$_{2}$\ & CH$_{2}$\ ``D'' \\
\tableline
Yield\tablenote{including uncertainty from background subtraction} 
             & 1.6 & 1.0 (0.7) & 1.4 (0.8) &  1.4 & 1.7 \\
Solid Angle: &     &           &           &      &     \\
\hspace{5mm}MC statistics& 0.4 & 0.4       & 0.4       & 0.4  & 0.4 \\
\hspace{5mm}R$_{\pi 2}\pm$3mm          & 0.5 & 0.5 & 0.5 & 0.5 & 0.5 \\
\hspace{5mm}${\pi 2}$ Area             & 0.3 & 0.3 & 0.3 & 0.3 & 0.3 \\
\hspace{5mm}hadronic losses            & 0.3 & 0.4 & 0.3 & 0.3 & 0.3 \\
\tableline
$\sqrt{\sum(\delta_{i})^2}$ & 1.8 & 1.3 (1.0) & 1.6 (1.2) & 1.6 & 1.9\\
\end{tabular}
\end{center}
\end{table}

\begin{table}
\caption{
Centre--of--mass absolute  differential cross sections at 141.15 MeV.  All 
the stated uncertainties are at
the 1$\sigma$ level.  The angle--dependent uncertainties and the normalization
uncertainty $\Delta$N are quadrature sums of the various
contributions listed in Table \protect\ref{tab:typerrs}.
The label $\Sigma$ indicates the {\em linear}\/ sum of the individual
normalization uncertainties. The uncertainty in the scatering angle
$\Delta\theta_{cms} = \pm$0.1$^{\circ}$.
}
\label{tab:dsg_results141}
\begin{center}
\begin{tabular}{lcc@{ $\pm$ }lc@{ $\pm$ }l} 
 \multicolumn{2}{l}{T$_{\pi}$ = 141.15 $\pm$ 0.6 MeV}  &
 \multicolumn{4}{c}{Absolute Differential Cross Sections [mb/sr]}  \\
\tableline
 Setup & $\theta^{\circ}_{\text{cms}}$ & 
 \multicolumn{2}{c}{$\frac{\text{d}\sigma}{\text{d}\Omega}$($\pi^{+}p$)}  &
 \multicolumn{2}{c}{$\frac{\text{d}\sigma}{\text{d}\Omega}$($\pi^{-}p$)}  \\
\tableline
             & & \multicolumn{2}{c}{[$\Delta$N = 1.1\% ($\Sigma$ = 2.3\%)]} &
                \multicolumn{2}{c}{[N/A]}\\
LH$_{2}$       & 25.3 & 15.17 & 0.45 & \multicolumn{2}{c}{-} \\
Single Arm     & 37.7 & 13.72 & 0.23 & \multicolumn{2}{c}{-} \\
               & 49.8 & 11.17 & 0.17 & \multicolumn{2}{c}{-} \\
               & 61.6 &  8.77 & 0.15 & \multicolumn{2}{c}{-} \\
               & 72.9 &  7.27 & 0.17 & \multicolumn{2}{c}{-} \\
               & 83.8 &  6.61 & 0.13 & \multicolumn{2}{c}{-} \\
\tableline
             & & \multicolumn{2}{c}{[$\Delta$N = 1.3\% ($\Sigma$ = 2.7\%)]} &
                 \multicolumn{2}{c}{[N/A]}  \\
LH$_{2}$            &  67.3 &  7.92 & 0.14 & \multicolumn{2}{c}{-} \\
Two Arm             &  78.4 &  6.71 & 0.12 & \multicolumn{2}{c}{-} \\
                    &  99.4 &  7.68 & 0.13 & \multicolumn{2}{c}{-} \\
                    & 118.6 & 11.76 & 0.18 & \multicolumn{2}{c}{-} \\
                    & 136.2 & 17.24 & 0.24 & \multicolumn{2}{c}{-} \\
                    & 152.7 & 21.63 & 0.29 & \multicolumn{2}{c}{-} \\
\tableline 
             & &  \multicolumn{2}{c}{[$\Delta$N = 1.3\% ($\Sigma$ = 2.6\%)]} &
                  \multicolumn{2}{c}{[$\Delta$N = 1.6\% ($\Sigma$ = 3.4\%)]} \\
LH$_{2}$            &  72.9 &  7.22  & 0.10  & 0.936 & 0.012 \\
Two Arm             &  89.1 &  6.69  & 0.08  & 0.695 & 0.010 \\
                    & 109.2 &  9.37  & 0.11  & 0.669 & 0.010 \\
                    & 127.6 & 14.59  & 0.15  & 0.944 & 0.014 \\
                    & 144.6 & 19.71  & 0.19  & 1.247 & 0.017 \\
                    & 160.7 & 23.26  & 0.23  & 1.426 & 0.020 \\
\tableline 
             &  & \multicolumn{2}{c}{[$\Delta$N = 1.3\% ($\Sigma$ = 2.6\%)]} &
                  \multicolumn{2}{c}{[$\Delta$N = 1.6\% ($\Sigma$ = 3.4\%)]} \\
2 mm CH$_{2}$       &  72.9  &  7.19 & 0.11 & 0.944 & 0.016 \\
Two Arm             &  89.1  &  6.69 & 0.11 & 0.686 & 0.015 \\
                    & 109.2  &  9.31 & 0.14 & 0.728 & 0.016 \\
                    & 127.6  & 14.53 & 0.19 & 0.973 & 0.020 \\
                    & 144.6  & 19.50 & 0.24 & 1.279 & 0.025 \\
                    & 160.7  & 23.87 & 0.28 & 1.509 & 0.030 \\
\end{tabular} 
\end{center}
\end{table}

\begin{table}
\caption{
Centre--of--mass absolute $\pi^+$p differential cross sections at 154.6 MeV. 
 }
\label{tab:dsg_results155}
\begin{center}
\begin{tabular}{lcc@{$ \pm$ }l} 
 \multicolumn{2}{l}{T$_{\pi}$ = 154.6 $\pm$ 0.7 MeV}  &
 \multicolumn{2}{c}{Absolute Differential Cross Sections [mb/sr]}  \\
\tableline
Setup & $\theta^{\circ}_{\text{cms}}$ & 
 \multicolumn{2}{c}{$\frac{\text{d}\sigma}{\text{d}\Omega}$($\pi^{+}p$)}  \\
\tableline
         &  &  \multicolumn{2}{c}{[$\Delta$N = 1.4\% ($\Sigma$ = 2.9\%)]} \\
LH$_{2}$            &  73.4  &  8.97  & 0.12 \\
Two Arm             &  89.6  &  7.81  & 0.10 \\
                    & 109.6  & 10.74  & 0.14 \\
                    & 128.0  & 16.66  & 0.20 \\
                    & 144.9  & 22.88  & 0.26 \\
                    & 160.8  & 27.28  & 0.32 \\
\end{tabular}
\end{center}
\end{table}

\begin{table}
\caption{
 Centre--of--mass absolute differential cross sections at 168.8 MeV.
 }
\label{tab:dsg_results169}
\begin{center}
\begin{tabular}{lcc@{ $\pm$ }lc@{ $\pm$ }l} 
 \multicolumn{2}{l}{T$_{\pi}$ = 168.8 $\pm$ 0.7 MeV}  &
 \multicolumn{4}{c}{Absolute Differential Cross Sections [mb/sr]}  \\
\tableline
 Setup & $\theta^{\circ}_{\text{cms}}$ & 
 \multicolumn{2}{c}{$\frac{\text{d}\sigma}{\text{d}\Omega}$($\pi^{+}p$)}  &
 \multicolumn{2}{c}{$\frac{\text{d}\sigma}{\text{d}\Omega}$($\pi^{-}p$)}  \\
\tableline
            & & \multicolumn{2}{c}{[$\Delta$N = 1.1\% ($\Sigma$ = 2.5\%)]} &
                \multicolumn{2}{c}{[N/A]} \\
LH$_{2}$     & 25.8 & 25.50 & 0.51 &  \multicolumn{2}{c}{-} \\
Single Arm   & 38.4 & 22.05 & 0.32 &  \multicolumn{2}{c}{-} \\
             & 50.6 & 17.04 & 0.24 &  \multicolumn{2}{c}{-} \\
             & 62.5 & 13.26 & 0.19 &  \multicolumn{2}{c}{-} \\
             & 73.9 & 10.34 & 0.16 &  \multicolumn{2}{c}{-} \\
             & 84.9 &  8.48 & 0.15 &  \multicolumn{2}{c}{-} \\
\tableline
           & & \multicolumn{2}{c}{[$\Delta$N = 1.3\% ($\Sigma$ = 2.7\%)]} &
                \multicolumn{2}{c}{[N/A]} \\
LH$_{2}$    &  68.3 & 11.42 & 0.18 & \multicolumn{2}{c}{-} \\
Two Arm     &  79.4 &  8.95 & 0.15 & \multicolumn{2}{c}{-} \\
            & 100.4 &  9.09 & 0.14 & \multicolumn{2}{c}{-} \\
            & 119.4 & 14.14 & 0.20 & \multicolumn{2}{c}{-} \\
            & 136.9 & 20.83 & 0.27 & \multicolumn{2}{c}{-} \\
            & 153.2 & 26.71 & 0.33 & \multicolumn{2}{c}{-} \\
\tableline 
           & &  \multicolumn{2}{c}{[$\Delta$N = 1.1\% ($\Sigma$ = 2.3\%)]} &
                \multicolumn{2}{c}{[$\Delta$N = 1.4\% ($\Sigma$ = 3.1\%)]} \\
LH$_{2}$    &  73.9 & 10.19  & 0.12  & 1.170 & 0.018 \\
Two Arm     &  90.2 &  8.46  & 0.08  & 0.846 & 0.013 \\
            & 110.1 & 11.15  & 0.10  & 0.960 & 0.015 \\
            & 128.4 & 17.26  & 0.14  & 1.475 & 0.021 \\
            & 145.2 & 23.90  & 0.19  & 2.069 & 0.028 \\
            & 161.0 & 28.82  & 0.23  & 2.430 & 0.033 \\
\tableline 
           & & \multicolumn{2}{c}{[$\Delta$N = 1.3\% ($\Sigma$ = 2.6\%)]} &
               \multicolumn{2}{c}{[$\Delta$N = 1.5\% ($\Sigma$ = 3.2\%)]} \\
2 mm CH$_{2}$ &  73.9  & 10.13 & 0.15 & 1.163 & 0.020 \\
Two Arm       &  90.2  &  8.57 & 0.14 & 0.884 & 0.018 \\
              &  110.1 & 11.21 & 0.18 & 0.999 & 0.021 \\
              &  128.4 & 17.78 & 0.25 & 1.468 & 0.029 \\
              &  145.2 & 23.83 & 0.32 & 2.091 & 0.037 \\
              &  161.0 & 28.93 & 0.37 & 2.516 & 0.043 \\
\end{tabular} 
\end{center}
\end{table}

\begin{table}
\caption{
 Centre--of--mass absolute  differential cross sections at 193.15 MeV.
As dicussed in Sec.~\protect\ref{sec:results},  the 
normalization uncertainties were increased to 2.5\% and 2.0\% for
 \mbox{$\pi^+$p}\/  and \mbox{$\pi^-$p}\/ scattering, respectively, to
 account for the systematic difference between the CH$_2$ and LH$_2$
 results at this one energy. To mimic the results at other energies, 
the linear sum uncertainty
 ($\Sigma$) was arbitrarily set to double the 1$\sigma$ uncertainty.
 }
\label{tab:dsg_results193}
\begin{center}
\begin{tabular}{lcc@{ $\pm$ }lc@{ $\pm$ }l} 
 \multicolumn{2}{l}{T$_{\pi}$ = 193.15 $\pm$ 0.7 MeV}  &
 \multicolumn{4}{c}{Absolute Differential Cross Sections [mb/sr]}  \\
\tableline
 Setup & $\theta^{\circ}_{\text{cms}}$ & 
 \multicolumn{2}{c}{$\frac{\text{d}\sigma}{\text{d}\Omega}$($\pi^{+}p$)}  &
 \multicolumn{2}{c}{$\frac{\text{d}\sigma}{\text{d}\Omega}$($\pi^{-}p$)}  \\
\tableline
       &  & \multicolumn{2}{c}{[$\Delta$N = 2.5\% ($\Sigma$ = 5.0\%)]} &  
            \multicolumn{2}{c}{[$\Delta$N = 2.0\% ($\Sigma$ = 4.0\%)]} \\
LH$_{2}$    &  74.8 & 10.03  & 0.12  & 1.101 & 0.018 \\
Two Arm     &  91.0 &  7.52  & 0.09  & 0.824 & 0.014 \\
            & 110.9 &  9.29  & 0.11  & 0.987 & 0.018 \\
            & 129.0 & 14.82  & 0.16  & 1.581 & 0.025 \\
            & 145.7 & 20.43  & 0.21  & 2.211 & 0.034 \\
            & 161.3 & 25.04  & 0.25  & 2.729 & 0.040 \\
\tableline 
  &  & \multicolumn{2}{c}{[$\Delta$N = 2.5\% ($\Sigma$ = 5.0\%)]} &
       \multicolumn{2}{c}{[$\Delta$N = 2.0\% ($\Sigma$ = 4.0\%)]} \\
2 mm CH$_{2}$ & 74.8  & 10.36 & 0.14 & 1.166 & 0.020 \\
Two Arm       & 91.0  &  7.91 & 0.13 & 0.835 & 0.019 \\
              & 110.9 &  9.86 & 0.15 & 1.001 & 0.023 \\
              & 129.0 & 15.51 & 0.22 & 1.691 & 0.034 \\
              & 145.7 & 21.73 & 0.28 & 2.342 & 0.042 \\
              & 161.3 & 26.15 & 0.33 & 2.776 & 0.051 \\
\end{tabular}
\end{center}
\end{table}

\begin{table}
\caption{ 
 Centre--of--mass absolute differential cross sections at 218.1 MeV.
\normalsize }
\label{tab:dsg_results218}
\begin{center}
\begin{tabular}{lcc@{ $\pm$ }lc@{ $\pm$ }l} 
 \multicolumn{2}{l}{T$_{\pi}$ = 218.1 $\pm$ 0.8 MeV}  &
 \multicolumn{4}{c}{Absolute Differential Cross Sections [mb/sr]}  \\
\tableline
 Setup & $\theta^{\circ}_{\text{cms}}$ & 
 \multicolumn{2}{c}{$\frac{\text{d}\sigma}{\text{d}\Omega}$($\pi^{+}p$)}  &
 \multicolumn{2}{c}{$\frac{\text{d}\sigma}{\text{d}\Omega}$($\pi^{-}p$)}  \\
\tableline
        & & \multicolumn{2}{c}{[$\Delta$N = 1.2\% ($\Sigma$ = 2.6\%)]} &
            \multicolumn{2}{c}{[N/A]}  \\
LH$_{2}$    & 26.5 & 26.16 & 0.46 & \multicolumn{2}{c}{-} \\
Single Arm  & 39.3 & 20.58 & 0.30 & \multicolumn{2}{c}{-} \\
            & 51.8 & 15.75 & 0.22 & \multicolumn{2}{c}{-} \\
            & 63.9 & 11.38 & 0.19 & \multicolumn{2}{c}{-} \\
            & 75.4 &  8.29 & 0.15 & \multicolumn{2}{c}{-} \\
            & 86.4 &  6.11 & 0.13 & \multicolumn{2}{c}{-} \\
\tableline
        & & \multicolumn{2}{c}{[$\Delta$N = 1.3\% ($\Sigma$ = 2.6\%)} &
            \multicolumn{2}{c}{[$\Delta$N = 1.5\% ($\Sigma$ = 3.0\%)} \\
LH$_{2}$    &  75.6 &  8.04  & 0.10  & 0.923 & 0.015 \\
Two Arm     &  91.9 &  5.62  & 0.06  & 0.663 & 0.011 \\
            & 111.8 &  6.72  & 0.07  & 0.847 & 0.014 \\
            & 129.7 & 10.59  & 0.11  & 1.353 & 0.020 \\
            & 146.2 & 14.85  & 0.14  & 1.948 & 0.027 \\
            & 161.5 & 18.41  & 0.17  & 2.395 & 0.033 \\
\tableline
       & & \multicolumn{2}{c}{[$\Delta$N = 1.3\% ($\Sigma$ = 2.8\%)]} &
           \multicolumn{2}{c}{[N/A]} \\
LH$_{2}$    &  69.9 &  9.49  & 0.13  & \multicolumn{2}{c}{-} \\
Two Arm     &  81.2 &  6.78  & 0.10  & \multicolumn{2}{c}{-} \\
            & 102.1 &  5.60  & 0.08 & \multicolumn{2}{c}{-} \\
            & 121.0 &  8.62  & 0.11  & \multicolumn{2}{c}{-} \\
            & 138.1 & 12.86  & 0.16  & \multicolumn{2}{c}{-} \\
            & 154.0 & 16.98  & 0.20  & \multicolumn{2}{c}{-} \\
\tableline
       &  & \multicolumn{2}{c}{[$\Delta$N = 1.5\% ($\Sigma$ = 3.1\%)]} &
            \multicolumn{2}{c}{[N/A]} \\
3.2 mm CH$_{2}$ &  69.9 &  9.61  & 0.12  & \multicolumn{2}{c}{-} \\
Two Arm         &  81.2 &  6.886 & 0.095 & \multicolumn{2}{c}{-} \\
                & 102.1 &  5.621 & 0.090 & \multicolumn{2}{c}{-} \\
                & 121.0 &  8.40  & 0.13  & \multicolumn{2}{c}{-} \\
                & 138.1 & 12.87  & 0.18  & \multicolumn{2}{c}{-} \\
                & 154.0 & 16.81  & 0.22  & \multicolumn{2}{c}{-} \\
\end{tabular}
\end{center}
\end{table}

\begin{table}
\caption{ 
 Centre--of--mass  absolute
 differential cross sections at 240.9 MeV.  The $\pi^{+}$\ point at 130.3$^0$
 has been deleted since the proton counter for this run 
 was found to be seriously misaligned. 
\normalsize }
\label{tab:dsg_results241}
\begin{center}
\begin{tabular}{lcc@{ $\pm$ }lc@{ $\pm$ }l} 
 \multicolumn{2}{c}{T$_{\pi}$ = 240.9 $\pm$ 0.9 MeV}  &
 \multicolumn{4}{c}{Absolute Differential Cross Sections [mb/sr]}  \\
\tableline
 Setup & $\theta^{\circ}_{\text{cms}}$ & 
 \multicolumn{2}{c}{$\frac{\text{d}\sigma}{\text{d}\Omega}$($\pi^{+}p$)}  &
 \multicolumn{2}{c}{$\frac{\text{d}\sigma}{\text{d}\Omega}$($\pi^{-}p$)}  \\
\tableline
       & & \multicolumn{2}{c}{[$\Delta$N = 1.3\% ($\Sigma$ = 2.8\%)]} &
           \multicolumn{2}{c}{[$\Delta$N = 1.4\% ($\Sigma$ = 2.9\%)]} \\
LH$_{2}$    &  76.3 &  6.35  & 0.08   & 0.822 & 0.013 \\
Two Arm     &  92.7 &  4.08  & 0.05   & 0.585 & 0.010 \\
            & 112.5 &  4.68  & 0.06   & 0.675 & 0.012 \\
            & 130.3 &   -    &  -     & 1.127 & 0.018 \\
            & 146.6 & 10.65  & 0.11   & 1.634 & 0.024 \\
            & 161.8 & 12.93  & 0.13   & 2.046 & 0.029 \\
\end{tabular}
\end{center}
\end{table}

\begin{table}
\caption{ 
 Centre--of--mass absolute differential cross sections 
 at 267.3 MeV. 
 }
\label{tab:dsg_results267}
\begin{center}
\begin{tabular}{lcc@{ $\pm$ }lc@{ $\pm$ }l} 
 \multicolumn{2}{l}{T$_{\pi}$ = 267.3 $\pm$ 0.9 MeV}  &
 \multicolumn{4}{c}{Absolute Differential Cross Sections [mb/sr]}  \\
\tableline
 Setup & $\theta^{\circ}_{\text{cms}}$ & 
 \multicolumn{2}{c}{$\frac{\text{d}\sigma}{\text{d}\Omega}$($\pi^{+}p$)}  &
 \multicolumn{2}{c}{$\frac{\text{d}\sigma}{\text{d}\Omega}$($\pi^{i}p$)}  \\
\tableline
     &   & \multicolumn{2}{c}{[$\Delta$N = 1.3\% ($\Sigma$ = 2.7\%)]} &
           \multicolumn{2}{c}{[$\Delta$N = 1.2\% ($\Sigma$ = 2.5\%)]} \\
LH$_{2}$     &  77.2 &  4.773 & 0.057 & 0.718 & 0.011 \\
Two Arm      &  93.5 &  2.793 & 0.036 & 0.508 & 0.010 \\
             & 113.3 &  3.025 & 0.041 & 0.575 & 0.011 \\
             & 131.0 &  4.812 & 0.059 & 0.835 & 0.016 \\
             & 147.1 &  6.959 & 0.081 & 1.263 & 0.023 \\
             & 162.1 &  8.526 & 0.098 & 1.542 & 0.031 \\
\tableline 
    &   & \multicolumn{2}{c}{[$\Delta$N = 1.3\% ($\Sigma$ = 2.8\%)]} &
          \multicolumn{2}{c}{[N/A]} \\
LH$_{2}$     &  71.4 &  5.779  & 0.075 &  \multicolumn{2}{c}{-} \\
Two Arm      &  82.8 &  3.875  & 0.058 &  \multicolumn{2}{c}{-} \\
             & 103.7 &  2.565  & 0.045 &  \multicolumn{2}{c}{-} \\
             & 122.3 &  3.862  & 0.062 &  \multicolumn{2}{c}{-} \\
             & 139.2 &  6.134  & 0.088 &  \multicolumn{2}{c}{-} \\
             & 154.7 &  7.93   & 0.11  &  \multicolumn{2}{c}{-} \\
\end{tabular}
\end{center}
\end{table}


\begin{references}
\bibitem{pinNewsLett} Many articles documenting these discrepancies can be
  found in the ``$\pi$N Newsletter'' series, volumes 1-15, published by 
UCLA and IEKP-Karlsruhe.
\bibitem{reginatubes} The transistorized tubes and
  bases were proprietary designs of the nuclear physics group at the
  University of Regina, Saskatchewan.
\bibitem{m11-joramhelp} C. Joram, Universit\"at Karlsruhe (private
  communication).  After noticing a similar effect, this group
  pointed out to us the references which explained the discrepancy.
\end{references}
\end{document}